\documentclass[onecolumn,11pt, journal]{IEEEtran}
\renewcommand*\footnoterule{}
\usepackage[round]{natbib}

\usepackage{amsmath}
\usepackage{amsfonts}
\usepackage{amsthm}   
\usepackage{mathrsfs}
\usepackage{tikz}
\usepackage[utf8]{inputenc}

\usepackage{epstopdf}
\usepackage{mathtools}
\usepackage{graphicx}
\usepackage{bbm}
\usepackage{enumerate}
\usepackage{epsf,verbatim,amssymb,array,cite,multicol,multirow}  
\usepackage{psfrag,bm,xspace}
\usepackage{hhline}
\usepackage{xstring}
\usepackage{ifthen}

\usepackage{booktabs} 

\usepackage{xparse}
\usepackage{physics}

\usepackage[linesnumbered,ruled,vlined,procnumbered]{algorithm2e}

\ifdefined \theorem 
\else
  \newtheorem{theorem}{Theorem}
\fi
\newtheorem{lem}{Lemma}

\ifdefined \lemma 
\else
\newtheorem{lemma}{Lemma}
\fi
\ifdefined \corollary 
\else
\newtheorem{corollary}{Corollary}
\fi

\newtheorem{fact}{Fact}

\ifdefined \proposition 
\else
\newtheorem{proposition}{Proposition}
\fi

\ifdefined \definition 
\else
\newtheorem{definition}{Definition}
\fi

\ifdefined \example 
\else
\newtheorem{example}{Example}
\fi

\ifdefined \remark 
\else
\newtheorem{remark}{Remark}
\fi

\newtheorem{innercustomgeneric}{\customgenericname}
\providecommand{\customgenericname}{}
\newcommand{\newcustomtheorem}[2]{%
  \newenvironment{#1}[1]
  {%
   \renewcommand\customgenericname{#2}%
   \renewcommand\theinnercustomgeneric{##1}%
   \innercustomgeneric
  }
  {\endinnercustomgeneric}
}

\newcustomtheorem{customthm}{Theorem}
\newcustomtheorem{customlemma}{Lemma}

\makeatletter
\def\old@comma{,}
\catcode`\,=13
\def,{%
  \ifmmode%
    \old@comma\discretionary{}{}{}%
  \else%
    \old@comma%
  \fi%
}
\makeatother

\newcommand\numberthis{\addtocounter{equation}{1}\tag{\theequation}}

\usepackage{xcolor}
\definecolor{darkblue}{rgb}{0.1,0.1,0.8}
\definecolor{DarkGreen}{rgb}{0,0.6,0}
\definecolor{brickred}{rgb}{0.8, 0.25, 0.33}
\definecolor{britishracinggreen}{rgb}{0.0, 0.26, 0.15}
\definecolor{calpolypomonagreen}{rgb}{0.12, 0.3, 0.17}
\definecolor{ao(english)}{rgb}{0.0, 0.5, 0.0}
	\definecolor{cadmiumgreen}{rgb}{0.0, 0.42, 0.24}
\definecolor{burgundy}{rgb}{0.5, 0.0, 0.13}
\usepackage{etoolbox}

\providetoggle{Blue_revision}
\settoggle{Blue_revision}{true}

\providetoggle{Track}
\settoggle{Track}{true}
\newcommand{\addv}[3]{%
	\iftoggle{Track}{%
    	\IfEqCase{#1}{%
       	 	{a}{\ifthenelse{\equal{#2}{ON}}{{\color{cadmiumgreen}#3}}{#3}}%
        	{b}{\ifthenelse{\equal{#2}{ON}}{{\color{brickred}#3}}{#3}}%
       		{c}{\ifthenelse{\equal{#2}{ON}}{{\color{burgundy}#3}}{#3}}%
       		{d}{\ifthenelse{\equal{#2}{ON}}{{\color{red}#3}}{#3}}
    	}[\PackageError{tree}{Undefined option to tree: #1}{}]%
	}{#3}%
}

 \usepackage{hyperref}

\hypersetup{
colorlinks=true, %
  pdfstartview={FitH},
    linkcolor=red,
    citecolor=blue,
    urlcolor={blue!80!black}
}

\usepackage{graphicx}

\newcounter{relctr} 
\everydisplay\expandafter{\the\everydisplay\setcounter{relctr}{0}} 

\AtBeginDocument{} 

\global\long\def\CC{\mathbb{C}}
\global\long\def\NN{\mathbb{N}}

\global\long\def\EE{\mathbb{E}}
\global\long\def\PP{\mathbb{P}}

\global\long\def\11{\mathbbm{1}}

\newcommand{\bfb}{\mathbf{b}}

\newcommand{\bfs}{\mathbf{s}}

\newcommand{\clJ}{\mathcal{J}}

\global\long\def\+{\oplus}

\newcommand\pmm{\{-1,1\}}

\newcommand{\prob}[1]{\PP\Big\{  #1 \Big\} }
\def\<{\langle}
\def\>{\rangle}

\DeclareMathOperator*{\sign}{sign}

\ifdefined \var 
  \renewcommand{\var}{\mathsf{var}}
\else
  \newcommand{\var}{\mathsf{var}}
\fi

\ifdefined \abs \else
 \newcommand{\abs}[1]{\lvert#1\rvert}
\fi

\ifdefined \norm \else
 \newcommand{\norm}[1]{\lVert#1\rVert}
\fi

\ifdefined \set 
  \renewcommand{\set}[1]{\left\{#1\right\}}
\else
  \newcommand{\set}[1]{\left\{#1\right\}}
\fi

\DeclareMathOperator*{\supp}{supp}

\newcommand*{\medcup}{\mathbin{\scalebox{1}{\ensuremath{\bigcup}}}}%
\usepackage{stackengine}
\def\deq{:=}

\DeclareMathOperator*{\tensor}{\otimes}

\providecommand{\tr}{tr}

\ifdefined \Tr 
  \renewcommand{\Tr}[1]{\tr \Big\{#1\Big\}}
\else
  \newcommand{\Tr}[1]{\tr \Big\{#1\Big\}}
\fi

 \def\id{I_d}

\def\gm{\emph{G}_M}

\def\qgs{{g}_{\bfs}}
\def\qYJ{\emph{F}_Y^{\mathcal{J}}}

\def\qfs{f_{\bfs}}
\def\qfests{\hat{f}_{\bfs}}

\def\qms{M_{\bfs}}

\def\qps{\sigma^\bfs}
\def\qfY{\emph{F}_{Y}}
\def\qfYest{{\hat{F}_Y}}

\def\qfsest{\hat{f}_{\bfs}}
\def\qfYestJ{{\qfYest}^{\mathcal{J}}}

\def\Msopt{\Lambda^{\bfs}}

\newcommand{\opt}{\optfont{opt}}
\def\Loss{L_{0-1}}

\def\GJ{G_{\mathcal{J}}}
\def\PiJhat{{\hat{\Pi}}^{\mathcal{J}}}

\def\L2{\mathcal{L}^2(\mathcal{X}^d, P_{X^d})}

\newcommand{\optfont}[1]{\mathsf{#1}}
\newcommand{\Popt}{\opt_{k}}

\def\E_mu{\mathcal{E}_{\mu}(\epsilon')}

\usepackage[nolist,nohyperlinks]{acronym}

\newacro{ptp}[PtP]{Point-to-Point}
\newacro{iid}[i.i.d.]{independently and identically distributed} 
\newacro{IID}[i.i.d.]{independently and identically distributed} 
\newacro{UFFS}[UFFS]{Unsupervised Fourier Feature Selection}
\newacro{SFFS}[SFFS]{Supervised Fourier Feature Selection}
\newacro{LS}[LS]{Laplacian Score}
\newacro{MAE}[MAE]{mean absolute error}
\newacro{MSE}[MSE]{mean square error}
\newacro{PAC}[PAC]{\textit{probably approximately correct}}
\newacro{VC}[VC]{Vapnik–Chervonenkis}
\newacro{ERM}[ERM]{Empirical Risk Minimization}
\newacro{SVM}[SVM]{support-vector machine}

\newacro{POVM}[POVM]{positive operator-valued measure}
\newacro{QLD}[QLD]{quantum low-degree algorithm}
\newacro{QNN}[QNN]{quantum neural networks}
\newacro{QP}[QP]{quantum perceptrons}
\newacro{SGD}[SGD]{stochastic gradient descent}
\newacro{QSGD}[QSGD]{quantum stochastic gradient descent}

\newacro{QC}[QC]{quantum computers}
\newacro{ML}[ML]{machine learning}
\newacro{QML}[QML]{quantum-enhanced machine learning}
\newacro{NISQ}[NISQ]{noisy intermediate-scale quantum}
\newacro{VQA}[VQA]{variational quantum algorithms}
\newacro{QSS}[QSS]{Quantum Shadow Sampling}

\settoggle{Track}{false}
\settoggle{Blue_revision}{false}

\newcommand{\addvc}[1]{\addv{c}{off}{#1}}

\newcommand{\addvf}[1]{\addv{b}{ON}{#1}}

\makeatletter
\def\footnoterule{\kern-3\p@
  \hrule \@width 6.5in \kern 2.6\p@} 
\makeatother
\begin{document}
\setlength{\baselineskip}{16pt}

\title{Learning $k$-qubit Quantum Operators via Pauli Decomposition}

\author{
\IEEEauthorblockN{ Mohsen Heidari \IEEEauthorrefmark{1} and  Wojciech Szpankowski \IEEEauthorrefmark{2}}\\
       \IEEEauthorrefmark{1}   Department of Computer Science, Indiana University, Bloomington \\
      \IEEEauthorrefmark{2} 
Department of Computer Science, Purdue University\\
\IEEEauthorrefmark{1}\tt mheidar@iu.edu,
 \IEEEauthorrefmark{2}\tt szpan@purdue.edu
   \thanks{This work was partially supported by the NSF Center for Science of Information (CSoI) Grant
CCF-0939370, and also by NSF Grants CCF-2006440, CCF-2007238, CCF-2211423, and Google Research Award.}   }
\IEEEoverridecommandlockouts
\maketitle

\begin{abstract}
Motivated by the limited qubit capacity of current quantum systems, we study the quantum sample complexity of  $k$-qubit quantum operators, i.e., operations applicable on only $k$ out of $d$ qubits. The problem is studied according to the quantum probably approximately correct (QPAC) model abiding by quantum mechanical laws such as no-cloning, state collapse, and measurement incompatibility. With the delicacy of quantum samples and the richness of quantum operations, one expects a significantly larger quantum sample complexity. 

This paper proves the contrary. We show that the quantum sample complexity of $k$-qubit quantum operations is  comparable to the classical sample complexity of their counterparts (juntas), at least when $\frac{k}{d}\ll 1$. This is surprising, especially since sample duplication is prohibited, and measurement incompatibility would lead to an exponentially larger sample complexity with standard methods.  Our approach is based  on the Pauli decomposition of quantum operators and a technique that we name Quantum Shadow Sampling (QSS) to reduce the sample complexity exponentially. The results are proved by developing (i) a connection between the learning loss and the Pauli decomposition;  (ii) a scalable QSS circuit for estimating the Pauli coefficients; and (iii) a quantum algorithm for learning $k$-qubit operators with sample complexity $O(\frac{k4^k}{\epsilon^2}\log d)$. 
\end{abstract}

\section{Introduction }
 Quantum-enhanced learning is one of the leading applications of \ac{QC} both for classical data  \citep{Giovannetti2008,Park2019,Lloyd2014,Schuld_2020} and inherently quantum samples 
 \citep{Carleo2017,massoli2021leap,Lu2018}.  However, current state-of-the-art \ac{QC}s have a limited qubit capacity of up to a few hundred qubits with infidelity. On the other hand, the dimension of quantum  systems in typical applications far exceeds the qubit capacity of \textit{near-term} \ac{QC}s. \addvf{Therefore, it is crucial to understand the  fundamental limits of near-term QCs for learning applications. Motivated by this observation, we study the learning capability of $k$-qubit operations in $d$-qubit systems, where $k$ is significantly smaller than $d$. Particularly, we characterize bounds on the quantum sample complexity of $k$-qubit systems and propose  a quantum learning algorithm achieving the minimum learning loss.  } 
 
 There are several quantum learning models such as state discrimination, quantum property testing, and quantum state classification (see Section \ref{subsec:related} for related works). For a comprehensive study, we consider a general formulation incorporating such models as special cases.  In classical settings, \ac{PAC}, developed by \citep{Kearns1994,Valiant1984}, is a concrete model to study fundamental limits such as sample complexity without any distributional or structural assumptions.
 In this work, we consider the quantum counterpart of this model, known as QPAC \citep{HeidariQuantum2021}. 

This learning model consists of a set of $n$ labeled qubits $(\rho_i, y_i)_{i=1}^n$ as the training samples. These samples might be classical or quantum originally. There is no structural assumption about the samples other than (1) being $d$-qubit states and (2) being generated \ac{iid} according to an unknown  but fixed probability distribution. The samples are processed by a \ac{QC} with a measurement at the end layer.  
We seek a procedure that takes the training set and tunes the quantum operations based on a library of choices (concept class). The objective is to  minimize the loss in predicting labels of the next unseen quantum states. Quantum sample complexity is, then, the minimum number of required samples to obtain a minimal loss.  

It is not difficult to see that this model subsumes several well-studied models. For example,  state discrimination is a special case in which $\rho_i$'s are identical and  are equal to one of two known possible states. Classical learning is also a special case in which $\rho_i$'s are pure states $\ketbra{x_i}$ with $x_i$ representing the classical samples.  \addvf{Therefore, QPAC is a stronger requirement than these special cases, as it is a distribution-free and state-free condition. Whereas PAC is only distribution-free, and state discrimination assumes certain prior structures.
In addition, more difficulties arise from the quantum nature of the problem. The quantum samples are irreversibly disturbed by the algorithm due to state collapse. Further, sample duplication is prohibited abiding by the \textit{no-cloning} principle.} 
 
  With the delicacy of quantum samples and
the richness of quantum operations, quantum sample complexity is  expected
to be significantly larger than the classical one. 
To see this, one natural approach for  learning $k$-qubit operations in $d$-qubit systems is via \textit{state tomography} with classical post proceeding. One first performs \textit{state tomography} on each sample to arrive at an approximate description of the quantum states and then performs classical learning algorithms on the stored density matrices. 
This approach requires $O(\frac{1}{\epsilon^2}2^{2d})$ identical copies of the samples \citep{Haah2016}. 
Compared to the classical setting, one considers learning of $k$-juntas which are  Boolean functions depending on $k$ out of $d$ inputs \citep{Mossel_ODonnell}. It is known that the sample complexity of  $k$-juntas is $O(\frac{k}{\epsilon^2}\log \frac{d}{k})$ \citep{ShalevShwartz2014}. \addvf{This observation implies that  the quantum sample complexity might be   exponentially larger.} 

\paragraph{Contributions:}
\addvf{In this paper, we prove the contrary and show that the quantum sample complexity of $k$-qubit operations scales logarithmically with $d$. Although not equal,  it is comparable to the classical sample complexity of $k$-juntas for small values of $k$.} More precisely, we prove in Theorem \ref{thm: Kjunta sample complexity} that the quantum sample complexity of $k$-qubit operations is $O(\frac{k4^k}{\epsilon^2}\log d)$. Furthermore, we  strengthen this existential result by designing a quantum algorithm achieving this bound (see Algorithm \ref{alg: QLD juntas}). 

Our approach is based on a Pauli decomposition of quantum operators and an estimation procedure called \ac{QSS}. We establish a connection between the learning loss and the Pauli decomposition of the induced operator of the training samples. We then develop a novel approach for estimating the Pauli coefficients of this induced operator. We argue that  naive empirical estimations require $O(\frac{(4d)^k}{\epsilon^2})$ quantum samples. Hence, they lead to an exponentially larger sample complexity than classical (See Section \ref{subsec:Pauli Estimation}). 
We address this issue and propose \ac{QSS} that reduces the quantum sample complexity to $O(\frac{k4^k}{\epsilon^2}\log d)$ that scales with the logarithm of the number of qubits (see Theorem \ref{thm:Pauli Estimation Shadow} and Section \ref{subsec:estimation circuit}). For that, we design an estimation circuit with $O(d)$ gate complexity. This design is  scalable as it consists of a parallel set of completely independent sub-circuits, each acting on a single qubit (see Figure \ref{fig:Shadow dqubit}). With this estimation, we develop our algorithm and prove that it  learns the $k$-qubit operators without any distributional or structural assumption (\textit{agnostic} QPAC). Lastly, in Section \ref{subse:numerical}, we verify our results with a numerical experiment for detecting maximally entangled from separable qubits.

 \subsection{Related Works}\label{subsec:related}
The literature in this area is broad. We only can give pointers to a few of the best-known and most relevant works. 

Quantum enhanced learning has been studied extensively for classical data \citep{Schuld_2014,Giovannetti2008,Park2019,Rebentrost2014,Lloyd2013,Lloyd2014} and for quantum data  in recent literature in the context of diverse applications, including condensed matter for phase-of-matter detection \citep{Carrasquilla2017,Broecker2017}, ground-state search \citep{Carleo2017,Broughton2020,Biamonte2017}, entanglement detection \citep{Ma2018,massoli2021leap,Lu2018,Hiesmayr2021,Chen2021a,Deng2017}, and other applications \citep{Kassal2011,McArdle2020,Hempel2018,Cao2019,HeidariAAAI2022,Bauer2020}.

There are several solutions and models for quantum learning.
In  \textit{state tomography}, the objective
is to find an approximate description of an unknown quantum
state $\rho$ using measurements on multiple copies of the state. This problem has been studied under various distance/fidelity measures \citep{ODonnell2016,ODonnell2017,Haah2016}. 
 \textit{State Certification} can be viewed as a quantum counterpart of property testing in which we would like to check where $\rho=\sigma$ or $\epsilon$ far away from it \citep{Badescu2019,Bubeck2020}. This is again done by measuring multiple identical copies of $\rho$. A survey on this topic is provided in \citep{Montanaro2016}. In \textit{state discrimination} we want to tell whether $\rho=\sigma_1$ or $\sigma_2$  \citep{Barnett2009,Gambs2008,Guta2010}. Another framework is \textit{quantum hypothesis testing} as surveyed in \citep{Audenaert2008}. An operational view of learning quantum states is introduced by \citep{Aaronson2007}. In this work, the training samples are \ac{iid} measurements. The objective is to approximate the acceptance probability $\tr{E\rho}$ for most measurement $E$. Another related work in this line is \citep{Cheng2015}, where an unknown measurement $E$ is to be learned from samples.  The training samples are $\set{(\rho_i, \tr{E\rho_i})}_{i=1}^n$, where $\rho_i$'s are \ac{iid} random quantum states. At first glance, this formulation seems similar to our problem. However, as a careful reader will recognize, $\rho_i$'s are pre-measured states. Contrary to this model, in our work, simultaneous access to pre-measured states and the measurement's outcomes are prohibited. Another distinction is that the probabilities $\tr{E\rho_i}$ are unknown in this paper. 
Another direction is based on the well-known work of \citet{Bshouty1998}. In this model, we measure identical copies of a
\textit{superposition} state to solve a classical PAC learning problem. This model is also different from QPAC in our paper, as the concept class in QPAC consists of quantum measurements rather than classical functions. Hence, QPAC is expected to subsume its model as well.   Other related works in this area are \citep{Arunachalam2017,Arunachalam2018,Kanade2018,Bernstein1997,Servedio2004}. Lastly, estimating the decomposition of an operator with respect to a set of elementary operators has been studied in \citep{Crawford2019,Peruzzo2014}.


\section{Model Formulation}

\noindent\textbf{Notations:}  For shorthand, denote $[d]$ as $\set{1,2,...,d}$. Also, for any  $\bfs \in \{0,1,2,3\}^d$, define $\supp(\bfs)\deq \set{\ell\in[d]: s_\ell\neq 0}$. For any $d\in \NN$, let $H_d$ be the Hilbert space of $d$-qubits. 
The identity operator on $H_d$ is denoted by $\id$. 
As usual, a quantum state is defined as a \textit{density operator}; that is a Hermitian, unit-trace, and non-negative linear operator.  
 A quantum measurement $\mathcal{M}$ is  a \ac{POVM} represented by a set of operators $\mathcal{M}:=\{M_v, v\in\mathcal{V}\}$, where $\mathcal{V}$ is theset of possible outcomes, $M_v\geq 0$ for any $v\in \mathcal{V}$, and $\sum_{v\in \mathcal{V}} M_v =\id.$  For an operator $A$, denote $\norm{A}_1=\tr{|A|}$ as the trace norm, and $\norm{A}_2=\sqrt{\tr{A^\dagger A}}$ as Hilbert–Schmidt norm. 
 %


\subsection{Quantum Learning Model}
Before presenting the main results, we formally define our quantum learning model. 
 In this model \citep{HeidariQuantum2021}, the objective is to  distinguish between multiple groups of unknown quantum states without prior knowledge about the states. Available is only a training set of quantum states with a classical label determining their group index. We seek an agnostic procedure that given enough samples learns the labeling law. The model in the binary case is defined more precisely as follows. 

Let ${\rho}_0$ and ${\rho}_1$ be two unknown quantum states denoting each of the possible states of an unknown physical system. We associate to each state a label $y\in \set{0,1}$. Let $p_0=1-p_1\in (0,1)$ be an unknown probability distribution on $\set{0,1}$. Each time, a sample $\rho$ is randomly generated where $\rho=\rho_0$ with probability $p_0$ and $\rho=\rho_1$ with probability $p_1$. The objective is to tell which of the two states is generated without knowing what $\rho_0, \rho_1$ and/or $(p_0, p_1)$ are.   
Available are only $n$ training samples $\set{({\rho}_{y_i}, y_i)}_{i=1}^n$, generated \ac{iid} according to $(p_0,p_1)$. 
   We seek a procedure that given the training samples constructs a quantum measurement to distinguish  between $\rho_0$ and $\rho_1$ with high accuracy. 

   A predictor is a quantum measurement that acts on the quantum state and outputs $\hat{y}\in \set{0,1}$ as the predicted label. 

   The accuracy of a prediction measurement $\mathcal{M}:=\set{M_0, M_1}$ is determined by randomly generating a test sample $({\rho}_{y_{test}}, y_{test})$.  Without revealing $y_{test}$, we measure ${\rho}_{y_{test}}$ to get $\hat{y}_{test}$. We use the 0-1 loss to measure the prediction error, that is $1\set{y_{test}\neq \hat{y}_{test}}$.   
   \addvf{Hence, from Born's rule, the (expected) loss is calculated as  $$L_{0-1}(\mathcal{M}) = p_0\tr{M_{1}{\rho}_{0}}+p_1\tr{M_{0}{\rho}_{1}},$$ 
   where the first and the second trace are the probability that erroneously $\hat{y}_{test}=1$ and  $\hat{y}_{test}=0$, respectively.}
It is assumed that $\mathcal{M}$ belongs to a collection $\mathcal{C}$ of choices as the \textit{concept class}. With this setup, a quantum learning algorithm is a process that selects a predictor $\mathcal{M}$ from $\mathcal{C}$, with the training samples as the input. We are interested in algorithms with guaranteed learning irrespective of $\rho_0, \rho_1, p_0$ and $p_1$.

\begin{definition}[QPAC] \label{def:quantum PAC}
A quantum learning algorithm QPAC learns a measurement class $\mathcal{C}$  if there exists a function $n_{\mathcal{C}}: (0,1)^2\rightarrow \NN$ such that for every $\epsilon, \delta \in [0,1]$ and given $n>n_{\mathcal{C}}(\epsilon,\delta)$  samples drawn \ac{iid} according to any probability distributions $(p_0, p_1)$ and from any unknown states $(\rho_0, \rho_1)$, the algorithm outputs, with probability of at least $(1-\delta)$, a measurement whose loss is less than $\inf_{\mathcal{M}\in \mathcal{C}} \Loss(\mathcal{M})+\epsilon$.\footnote{\addvf{Naturally, we are interested in efficient learning with   $n_{\mathcal{C}}$ being at most polynomial in $\epsilon, \delta$ and $d$.}}
\end{definition}

Consequently, the quantum sample complexity of a concept class $\mathcal{C}$ is the minimum of $n_{\mathcal{C}}$ for which there exists a QPAC learning algorithm. 
The focus of this study is on $k$-qubit operators that are formally defined below. 

\begin{definition}[$k$-qubit Operators] \label{def:k-qubit measurement}
An operator $A$ on $H_d$ is said to be a $k$-qubit operator, if there exists a coordinate subset $\mathcal{J}\subset [d]$ with $|\mathcal{J}|\leq k$ such that  $A=\tilde{A}_{\mathcal{J}}\tensor I_{[d]\backslash \mathcal{J}}$, where $\tilde{A}$ is an operator on the subsystem corresponding to the coordinates $\mathcal{J}$ 
and $I_{[d]\backslash \mathcal{J}}$ is the identity operator on the residual sub-system.
\end{definition}

\addvf{Classical counter parts of $k$-qubit operators are $k$-junta Boolean functions  \citep{mossel2004learning}. $k$-qubit operators subsumes $k$-juntas. They are significantly richer  than their classical counterpart.  While there are ${d \choose k} 2^{2^k}$ juntas; $k$-qubit operators are infinite. The input dimension for a $k$-junta is $k$; while that of a $k$-qubit operator is $2^k$. One can learn $k$-juntas by performing a brute-force  exhaustive search over all $k$-juntas  and finding the one minimizing the empirical loss. However, the learning task becomes more difficult in the quantum settings as there are infinitely many $k$-qubit circuits and sample duplication is prohibited. Therefore,  with the richness of quantum concept classes and the
the fragility of quantum samples, one wonders whether quantum learning is harder.
In the next section, we show it is not, but it requires looking at the problem from a
different angle.}

\section{Main Results}
Our first main contribution is the following theorem that is proved in Section \ref{subsec:estimation circuit}.


 \begin{theorem}\label{thm: Kjunta sample complexity}
 There exists a quantum algorithm  that QPAC learns $k$-qubit operators with an error up to $$
 \opt_{k}+\order{\sqrt{ \frac{4^k}{n}\log(\frac{d^k 4^k}{\delta(k-1)!})}},$$
 where $\opt_k$ is the minimum loss of the concept class, and this is achieved by Algorithm \ref{alg: QLD juntas}. \end{theorem}

With this result, for small $\frac{k}{d}$, the quantum sample complexity of $k$-qubit operators  is simplified to $O(\frac{k4^k}{\epsilon^2}\log \frac{d}{\delta})$ which grows with logarithm of $d$, the number of qubits. 

Next, we study a lower bound on the quantum sample complexity.
Given that QPAC subsumes PAC and that $k$-juntas are special cases of $k$-qubit operations, the quantum sample complexity is bounded from below by the classical one. Hence, from \ac{VC} theory for $k$-juntas, \citep{ShalevShwartz2014}, we obtain the following lower bound. 
\begin{proposition}
The quantum sample complexity of $k$-qubit operations is $\Omega(\frac{1}{\epsilon^2}(k\log \frac{2d}{k} +\log\frac{1}{\delta}))$.
\end{proposition}
This result and Theorem \ref{thm: Kjunta sample complexity} suggest that quantum sample complexity is of the same order as the classical one at least for small values of $k$ compared to $d$. Hence, though QPAC is a more difficult problem and low-width quantum circuits are much richer than classical juntas, yet the quantum sample complexity grows similarly for small $k$'s. Whether the same holds for larger values of $k$ is yet to be determined. In Section \ref{subsec:Pauli Estimation}, we argue that primitive empirical estimation methods are not efficient in QPAC and that one needs a more sophisticated approach as in Algorithm \ref{alg: QLD juntas}. Before that, we present an overview of the Pauli decomposition and study its connection to learning loss.



\subsection{Pauli Decomposition}\label{subsec:QFoureir}
Our approach relies on the Pauli decomposition of quantum operators \citep{Montanaro2008}.   We start with a brief overview of this decomposition. Then, we analyze the connection between the $0-1$ loss and the Pauli coefficients. 

The Pauli operators  with the identity are denoted as $\{\sigma^0, \sigma^1, \sigma^2, \sigma^3\}$ with $\sigma^0=I_2$ and 
 \begin{align*}
    \sigma^1 = \begin{pmatrix}
 0 & 1\\ 1 & 0
 \end{pmatrix},
 \quad  \sigma^2 =  \begin{pmatrix}
 0 & -i\\ i & 0
 \end{pmatrix},
\quad  \sigma^3 =  \begin{pmatrix}
 1 & 0\\ 0 & -1
 \end{pmatrix}.
 \end{align*}  
 Define the Pauli tensor products as
\begin{align}\label{eq:Pauly for s}
\sigma^{\bfs}:= \sigma^{s_1}\tensor \sigma^{s_2}\tensor \cdots \tensor \sigma^{s_d}, \qquad \forall \bfs\in \{0,1,2,3\}^d.
\end{align}
%
\begin{fact}[Pauli Decomposition]\label{fact:quantum Fourier}
Any bounded  operator $A$ on $H_d$ is uniquely decomposed as   $$A = \sum_{\bfs \in \{0,1,2,3\}^d} a_{\bfs} ~\sigma^{\bfs},$$ where $a_\bfs\in \CC$ are the Pauli coefficients of $A$ {and are given by $a_\bfs=\frac{1}{2^d}\tr\big\{A\qps\big\}$.}\footnote{The factor $2^d$ is because $\tr{\qps\qps}=\tr{\id}=2^d$.} 
\end{fact}

An immediate consequence of this decomposition is the following identity for any pair of operators on $H_d$:
\begin{align}\label{eq:Fourier trace}
\tr{AB} = 2^d \sum_{\bfs} a_{\bfs}b_{\bfs},
\end{align}
where $a_{\bfs}$ and $b_{\bfs}$ are the Pauli coefficients of $A$ and $B$. 
Next, we present the connection between the Pauli coefficients and the learning loss. 

\begin{lem}\label{thm:mislabeling prob}
Let $\rho_{XY}=p_0 \rho_0 \tensor \ketbra{0}+p_1 \rho_1 \tensor \ketbra{1}$ denote the average state of the training samples.  Then, the loss of any measurement $\mathcal{M}:=\{M_0, M_1\}$ decomposes as 
$$\Loss(\mathcal{M})  = \frac{1}{2}- 2^{d-1}\sum_{\bfs} \qgs \qfs,$$ where $\qgs$ and $\qfs$ are the Pauli coefficients of $\gm:=M_1-M_0$ and $\qfY :=   -\sqrt{\rho_{XY}}( \id \tensor \sigma^3) \sqrt{\rho_{XY}}$, respectively.
\end{lem}
\begin{proof}
Given $M_1=I_d-M_0$, the  loss can be written as 
\begin{align*}
\Loss(\mathcal{M}) &= p_1\tr{M_0 {\rho}_1} + p_0\tr{M_1 {\rho}_0}\\\numberthis \label{eq:mislabeling 1}
&=-\sum_{y}p_{y} (-1)^y\tr{M_0 {\rho}_y} + p_0.
\end{align*}
Observe that $(\id\tensor \sigma^3)\rho_{XY} = \sum_{y}p_{y}(-1)^y {\rho}_y\tensor \ketbra{y}.$ 
Then, from the definition of $\qfY$ and $\gm$, we have that
\begin{align*}
\tr{\gm\qfY} &= \tr{\qfY}-2\tr{M_0\qfY}\\
&= -\tr{(\id\tensor \sigma^3)\rho_{XY}}-2\tr{M_0\qfY}\\
&= -\EE_{Y}[(-1)^Y] + 2\sum_{y} p_{y} (-1)^y\tr{M_0 {\rho}_y}\\
&\stackrel{(a)}{=}-\EE[(-1)^Y] + 2\Big(p_0 - \Loss(\mathcal{M}) \Big)\\
&= 2p_1-1 + 2\Big(p_0 - \Loss(\mathcal{M}) \Big)\\
&= 1 - 2 \Loss(\mathcal{M}), 
\end{align*}
where (a) follows from \eqref{eq:mislabeling 1}.
Hence,  $$\Loss(\mathcal{M}) = \frac{1}{2}-\frac{1}{2}\tr{\gm\qfY}.$$ Hence, the proof is complete, because from \eqref{eq:Fourier trace}, the trace term above equals to $2^d\sum_{\bfs} \qgs \qfs$.
\end{proof}
We note that $\qfY$ is  viewed as the induced operator representing the labeled samples. In agnostic settings,  $\qfY$ is unknown as the states and the probabilities are unknown. We design our learning algorithm by  estimating the Pauli coefficient of $\qfY$ instead.

\subsection{Estimating the Pauli Coefficients}\label{subsec:Pauli Estimation}
In light of the previous section, the main idea behind the proposed algorithm is to estimate a subset of the Pauli coefficients of the auxiliary operator $\qfY$.
In the classical setting, estimating the Fourier coefficients is easily done by empirical averaging. In quantum, each coefficient $\qfs$ is indeed an observable acting on the samples' quantum state. The issue is that these observables are \textit{incompatible} and, thus, are not simultaneously measurable.  
\begin{example}
In a single qubit system, the Pauli coefficients corresponding to $\sigma^1$ and $\sigma^2$ are incompatible as they do not \textit{commute} with each other. Indeed, they are \textit{mutually unbiased observables}. Hence, independent samples are needed for estimating each coefficient.
\end{example}

The incompatibility and no-cloning make the estimation process more challenging than the classical one. In this section, we discuss the estimation process and derive bounds on the square loss. In the next section, we discuss the construction of a predictor from the estimated coefficients. 


We start with estimating a single Pauli coefficient $\qfs$. For that we consider the POVM  $\qms := \{\Msopt_1, \Msopt_{-1} \}$ with outcomes in $\pmm$ and operators 
\begin{align}\label{eq:Ms operators}
\Msopt_{1} \deq \qps_+,\qquad \Lambda^\bfs_{-1} \deq \qps_-,
\end{align}
where $\qps$ is the Pauli operator corresponding to $\bfs\in \set{0,1,2,3}^d$ as in \eqref{eq:Pauly for s}. Moreover, $\qps_+$ and $\qps_-$ are the positive and negative part of $\qps$ (such that $\qps=\qps_+ - \qps_-$), constructed through the spectral decomposition of $\qps$.  With these definitions, $\qfs$ is estimated by measuring each sample with  $\qms$. Note that we cannot use all the samples for estimating one coefficient; because the samples will be inaccessible as they collapse by the measurements. Suppose, we only use $m<n$ samples.   Let $Z_i\in \pmm$ be the output of $\qms$ on the $i$th sample $(\rho_i, y_i), i=1,2,3..., m$. Then, the estimation is computed as  
\begin{equation}\label{eq:fs hat}
    \qfests = \frac{-1}{m2^d}\sum_{i=1}^m (-1)^{y_i} Z_i.
\end{equation}

 From Born's rule, $Z_i$ is a binary random variable with bias $\tr{\qps_+ \rho_i}$. Hence, $\qfests$ itself is random and it is not difficult to check that  $\EE[\qfests]=\qfs$.  Therefore, using standard concentration inequalities, we can show that, for any $\delta \in [0,1]$, with probability $(1-\delta)$, the estimation error is bounded as: 
\begin{align}\label{eq:naive single}
 |\qfsest - \qfs| = 2^{-d}\order{\sqrt{\frac{1}{m}\log\frac{1}{\delta}}}.
 \end{align} 
{Note that $2^{-d}$ is due to the normalization of $\qfs$ as in Fact \ref{fact:quantum Fourier}  and $\qfsest$ as in \eqref{eq:fs hat}.}
 
 For learning $k$-qubit measurements, all the Pauli coefficients $\qps$  with $|\supp(\bfs)|\leq k$ need to be estimated. Let $K$ be the number of such coefficients. Given that $k\leq d/2$, we bound $K$ as 
\begin{align}\label{eq:K}
K \leq \sum_{\ell=0}^k {d \choose \ell} 4^{\ell} \leq 1+ k {d \choose k} 4^k = 1+ \frac{d^{k}}{(k-1)!}4^k.
\end{align}
Given the incompatibility of the related observables, with a naive strategy, one would partition the total $n$ samples into several equal-size groups one for each coefficient. Hence, with this approach and \eqref{eq:naive single}, the estimation loss satisfies
\begin{align*}
|\qfsest - \qfs| =  \order{2^{-d}\sqrt{\frac{K}{n}\log(1/\delta)}},
\end{align*}
for all $\bfs$, with $|\supp(\bfs)|\leq k$. In what follows, we propose an approach to exponentially reduce the estimation error.

\begin{theorem}\label{thm:Pauli Estimation Shadow}
Given any  $\bfs_1, \bfs_2, \cdots, \bfs_K \in \set{0,1,2,3}^d$, there exists an algorithm that, given $n$ training samples, estimates the corresponding Pauli coefficients of $\qfY$ with an error bounded with probability at least $(1-\delta)$ as 
\begin{align*}
\sup_{j\in [K]}|\hat{f}_{\bfs_j} - f_{\bfs_j}| =  \order{2^{-d}\sqrt{\frac{1}{n}\log(\frac{K}{\delta})}}.
\end{align*}

\end{theorem}

For our case, $\bfs_j$'s are all $\bfs$ with $|\supp(\bfs)|\leq k$, and $K$ is as in \eqref{eq:K}. Hence, compared to the naive strategy with a fresh copy for each coefficient, we get exponential improvements.   
\subsubsection{Quantum Shadow Sampling}
  Our approach  is inspired by Shadow Tomography \citep{Aaronson2018,Huang2020}, where repeated measurements obtain an approximate description of an unknown quantum state from its exact copies. In view of the no-cloning, in our work, we propose an alternate approach called \ac{QSS} that takes a single quantum state and can generates multiple samples called shadows. This is a one-shot procedure that applies to each sample $(\rho_i, y_i), i\in [n]$ and is explained below: 

 First, we generate a  unitary operator $U_i$ randomly and uniformly from the space of all unitary operators on $d$ qubits. We rotate $\rho_i$ by applying $U_i$ resulting the state $U_i^\dagger \rho_i U_i$. Then, we measure the rotated state along the computational basis $\set{\ketbra{b}, b\in \{0,1\}^d}$. From Born’s rule the probability of getting  the output $b_i\in \set{0,1}^d$ is $P_{b_i}= \matrixelement{b_i}{U_i^{\dagger}\rho_i U_i}{b_i}$. 
At the next step, given each output $b_i\in \set{0, 1}^d$, the state $\omega_i = U_i\ketbra{b_i}U_i^\dagger$ is prepared. Hence, with $\rho_i$ we obtain the state $\omega_i$ with probability $P_{b_i}$. 

Define the following mapping on any operator $B$ on $H_d$:
\begin{align}\label{eq:Gamma}
\Gamma[B]:=\EE_{U}\Big[\sum_{b\in \set{0,1}^d} \matrixelement{b}{U^{\dagger}B U}{b}~ U\ketbra{b}U^\dagger\Big].
\end{align}
Note that $\Gamma$ is a linear mapping on the space of density operators with its inverse denoted as $\Gamma^{-1}$. Moreover, observe that $\Gamma[\rho_i]$ equals to the expectation $\EE[\omega_i]$ over the measurement randomness ($P_b$) and the choices of unitary $U_i$.

At our last step, we apply $\Gamma^{-1}$ on $\omega_i$ resulting in the following state 
\begin{align*}
\hat{\rho}_i:=\Gamma^{-1}\big[U_i\ketbra{b_i}U_i^\dagger\big].
\end{align*}

\begin{figure}[tbp]
\centering
\includegraphics[width=0.45\textwidth]{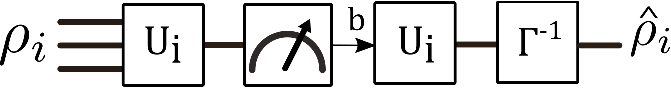}
\caption{The process for estimating the Pauli coefficient $\qfs$. Once $\hat{\rho}_i$ is generated from the $i$th sample, we calculate $\tr{\hat{\rho}_i \qps}(-1)^{y_i}$. Then, the estimate $\qfsest$ is calculated by computing the empirical average over all the samples as in \eqref{eq:pauli estimate }.   }
\label{fig:Estimation}
\end{figure}

 Repeating this process for all samples, we obtain the shadow samples $\hat{\rho}_i, i\in [n]$. This process is demonstrated in Figure \ref{fig:Estimation}. 
 
 \subsection{Proof of Theorem \ref{thm:Pauli Estimation Shadow}}
 After applying QSS on the entire training samples, we estimate each $f_{\bfs_j}$ by computing 
\begin{align}\label{eq:pauli estimate }
\hat{f}_{\bfs_j} = \frac{1}{n}2^{-d}\sum_{i=1}^n \tr{\hat{\rho}_i \sigma^{\bfs_j}}(-1)^{y_i}, 
\end{align}
for all $j\in[K]$. 
 We proceed with the following lemmas for the analysis.

\begin{lem}\label{lem:hat rho is unbiased}
$\hat{\rho}_i$ is an unbiased estimate of $\rho_i$, that is $\EE_{U,b}[\hat{\rho}_i]=\rho_i$. 
\end{lem}
\begin{proof}
By linearity of $\Gamma^{-1}$, taking the expectation of $\hat{\rho}_i$ over the choice of $U_i$ and the randomness of $b_i$ gives
\begin{align*}
\EE_{\sim (U_i, b_i)}[\hat{\rho}_i] &= \Gamma^{-1}\big[\EE[U_i\ketbra{b_i}U_i^\dagger]\big].
\end{align*}
The expectation term equals to
\begin{align*}
\EE\Big[U_i\ketbra{b_i}U_i^\dagger \Big] &= \EE_{U}\Big[\sum_{\bfb} \matrixelement{\bfb}{U^{\dagger}\rho_i U}{\bfb}~ U\ketbra{\bfb}U^\dagger\Big]\\
&=\Gamma[\rho_i],
\end{align*}
where the last equality is from \eqref{eq:Gamma}. 
\end{proof}

\begin{lem}\label{lem:Pauli estimation}
The estimation $\qfsest$ is unbiased, that is $\EE[\qfsest]=\qfs$, where the expectation is taken over all sources of randomness including the sample distribution.
\end{lem}
\begin{proof}
By taking the expectation, from Lemma \ref{lem:hat rho is unbiased} we obtain the following chain of equalities:
\begin{align*}
\EE[\qfsest] &= 2^{-d}\EE[\tr{\hat{\rho}_1 \qps}(-1)^{Y_1}]\\
& = 2^{-d}\EE_{\sim (\rho_1, Y_1)}[ \tr{\EE[\hat{\rho}_1| \rho_1] \qps}(-1)^{Y_1}]\\
&=2^{-d}\EE_{\sim (\rho_1, Y_1)}[ \tr{\rho_1 \qps}(-1)^{Y_1}]\\
&=2^{-d}\tr{\qfY\qps}=\qfs,
\end{align*}
where we used the definition of $\qfs$ in Lemma \ref{thm:mislabeling prob}.
\end{proof}

Lastly, with Lemma \ref{lem:Pauli estimation}, we apply the Chernoff inequality: 
\begin{align*}
\prob{\max_{j\in [K]} |\hat{f}_{\bfs_j} - f_{\bfs_j}| \geq \epsilon2^{-d} }\leq 2 K ~\text{exp}\Big\{-\frac{n\epsilon^2}{2}\Big\}.
\end{align*}
Equating the right-hand side to $\delta$, we obtain the following bound on the estimation error: 
\begin{align}\label{eq:Pauli estimation error}
\max_{j\in [K]} |\hat{f}_{\bfs_j} - f_{\bfs_j}|  &= \order{2^{-d}\sqrt{\frac{1}{n}\log(\frac{K}{\delta})}}.
\end{align}
 With this inequality, we establish Theorem \ref{thm:Pauli Estimation Shadow}.

\subsection{Creating the Predictor}
Next, we describe the construction of a predictor using the estimated Pauli coefficients. Let $\mathcal{J}\subseteq [d]$ be the coordinate of a subsystem with $k$ qubits.  Define $$\qYJ:=\sum_{\bfs: \supp(\bfs) \subseteq \mathcal{J}} \qfs \qps,$$
where $\supp(\bfs):= \set{\ell\in[d]: s_\ell\neq 0}$ for any $\bfs \in \{0,1,2,3\}^d$. Define the estimate of this operator as
\begin{align*}
\qfYestJ:=\sum_{\bfs: \supp(\bfs) \subseteq \mathcal{J}} \qfsest \qps,
\end{align*}
where $\qfsest$'s are the estimated Pauli coefficients.  This operator has a spectral decomposition of the form 
\begin{align*}
\qfYestJ =  \sum_{i} \lambda_i \ketbra{\phi_i}.
\end{align*}
Let $\PiJhat$ be the projection onto the subspace spanned by eigenstates with positive eigenvalues, i.e., 
\begin{align}\label{eq:Pi k}
  \PiJhat:=\sum_{i: \lambda_i>0} \ketbra{\phi_i}.
  \end{align}  
 Then, we create our predictor as the measurement $\hat{\mathcal{M}}_\clJ := \{\PiJhat, \id-\PiJhat\}$. 
In what follows, we study the learning loss of  $\hat{\mathcal{M}}_\clJ$. We show that if $\clJ$ is chosen appropriately, then the loss of $\hat{\mathcal{M}}_\clJ$ is close to the optimal value $\opt_k$. For that, we present the following theorem. 

   \begin{theorem}\label{thm:loss M_k}
Let $\mathcal{J}^*$ be the subset maximizing $\norm{\qYJ}_1$ among all $k$-element subsets. Let $\hat{\mathcal{M}}_{\mathcal{J}^*}=\{\hat{\Pi}^{\mathcal{J}^*}, I-\hat{\Pi}^{\mathcal{J}^*}\}$ be the measurement with the projection $\hat{\Pi}^{\mathcal{J}^*}$ given in \eqref{eq:Pi k} but with $\clJ=\clJ^*$.  Then,
\begin{align}\label{eq: 0-1 loss Pauli}
\Loss(\hat{\mathcal{M}}_{\clJ^*}) & \leq \Popt + 4\sqrt{2^d}\norm{\qYJ- \qfYestJ}_2
\end{align}
where $\opt_{k}$ is the minimum loss among all $k$-qubit operations, and $\qfsest$ is the estimation of $\qfs$.
\end{theorem}
This theorem implies an interesting connection between the QPAC learnability of a predictor and its Pauli decomposition. Moreover, it implies that the square loss is a suitable loss function for estimating the Pauli operators. 
We note that the factor $2^d$ is not problematic as it appears simply because of the way the Pauli coefficients are defined. 


\paragraph{Proof Sketch of Theorem \ref{thm:loss M_k}:}\label{subsec:proof:thm:loss Mk}
The proof of this theorem is involved. Here,  we only explain the sketch of its proof  by presenting the following key lemmas with their proof in Appendix \ref{sec:proofs}. The first lemma characterizes $\opt_{k}$ and the second lemma is the key connection to Pauli estimations in our analysis.

   \begin{lem}\label{lem: k junta opt}
If $\opt_{k}$ is the minimum loss among the class of all $k$-qubit measurements, then 
\begin{align*}
\opt_{k} = \frac{1}{2}-\frac{1}{2}\max_{\mathcal{J}\subset [d]: |\mathcal{J}|= k} \norm\big{\qYJ}_{1},
 \end{align*} 
 where $\norm{\cdot}_1$ is the trace norm. 
 \end{lem}



   \begin{lem}\label{lem:norm2 sign}
Let $\mathcal{M}_{\clJ}=\{\PiJhat, I-\PiJhat\}$ be the measurement with the projection $\PiJhat$ given in \eqref{eq:Pi k}.   Then,
\begin{align}\nonumber
\Loss(\mathcal{M}_{\mathcal{J}})&\leq \frac{1}{2}\Big(1-\norm{\qYJ}_{1}\Big) +U\Big(\sqrt{2^d}\norm{\qYJ- \qfYestJ}_2\Big), 
\end{align}
where $U(x)=x^3+\frac{3}{2}x^2+\frac{3}{2}x$, for all $x\geq 0$.
\end{lem}
With these lemmas, Theorem \ref{thm:loss M_k} immediately follows. 
Let  $\mathcal{J}^*$ be the  coordinate as in Theorem \ref{thm:loss M_k}.  Then, with  Lemma \ref{lem: k junta opt} and \ref{lem:norm2 sign}, and the fact that $U(x)\leq 4 x$ for $x\leq 1$, we have that
\begin{align*}
\Loss(\mathcal{M}_{\mathcal{J}^*})&\leq \opt_k+4 \sqrt{2^d}\norm{\qYJ- \qfYestJ}_2.
\end{align*}
With that, we establish the theorem.  
\subsection{Algorithm and Proof of the Main Theorem}\label{subsec:estimation circuit}
So far, we discussed the estimation of the Pauli coefficients and the construction of the predictor. The estimation process in Section \ref{subsec:Pauli Estimation} in its current form may not be applicable when $d$ is  large. It is not clear how to create $\Gamma^{-1}$ and $U_i$ in Figure \ref{fig:Estimation}. In this section, we characterize a closed-form expression for $\Gamma^{-1}$ and present an implementation of it with a scalable circuit.

Consider $\Gamma[\rho_i]$ as in \eqref{eq:Gamma} for a single qubit system ($d=1$). Instead of ranging over all unitary operators, we choose $U$ from the following set with equal probabilities: 
\begin{align*}
 U\in \set{ I, H, S^\dagger H},
 \end{align*} 
 where $H$ is the Hadamard and $S=\sqrt{\sigma^3}$. With this set, the state is measured either along the computational basis, $X$-basis, or the $Y$-basis.   Let
\begin{align*}
 \Gamma_0[\rho]:=\hspace{-10pt}\sum_{U\in \set{ I, H, S^\dagger H}}\sum_{b\in \set{0,1}} \frac{1}{3}\matrixelement{b}{U^{\dagger}\rho U}{b}~ U\ketbra{b}U^\dagger.
 \end{align*} 
 It is not difficult to check that $\Gamma_0$ has an inverse and that Lemma \ref{lem:hat rho is unbiased} still holds as the above set is tomographically complete. 
 
\begin{figure}[tbp]
\centering
\includegraphics[width=0.45\textwidth]{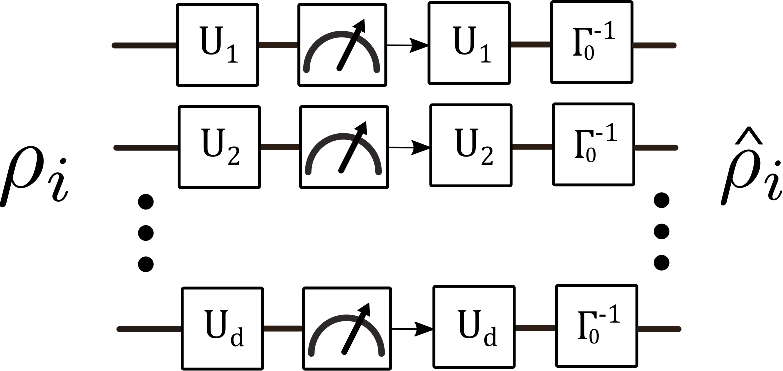}
\caption{Scalable implementation of the estimation process in Figure \ref{fig:Estimation}.  For each sample, $U_1, U_2, \cdots, U_d$ are selected randomly and independently from $\set{I,H, S^\dagger H}$. The circuits operate independently on each corresponding qubit implying a $O(d)$ gate complexity. 
   }
\label{fig:Shadow dqubit}
\end{figure}

 For general $d$-qubit systems, we apply the single-qubit process to each of the $d$ qubits independently (see Figure \ref{fig:Shadow dqubit}).  
In the following, we show that this circuit gives an unbiased estimate of $\rho$, even though $\rho$ could be an entangled state! 
\begin{lem}
Let $\hat{\rho}_i$ be the operation's output in Figure \ref{fig:Shadow dqubit}. Then, $\EE[\hat{\rho}_i]=\rho_i$. 
\end{lem}
\begin{proof}
Consider the Pauli decomposition of $\rho_i = \sum_{\bfs} \alpha_s \qps$. As $\Gamma_0^{-1}$ is a linear mapping, the circuit in Figure \ref{fig:Shadow dqubit} is also linear. Let $\Psi$ represent this operation. Then $\hat{\rho}_i=\Psi[\rho_i]$. The linearity implies that  $\Psi[\rho_i] = \sum_{\bfs} \alpha_s \Psi[\qps]$. Since, $\qps$ is in tensor product and $\Psi$ operates on each qubit independently, then $\Psi[\qps] = \tensor_{j=1}^d \Psi_j[\sigma^{s_j}]$, where $\Psi_j$ is the $j$th wire on Figure \ref{fig:Shadow dqubit}. As a result, 
\begin{align*}
\EE[\hat{\rho}_i] = \sum_{\bfs} \alpha_s \bigotimes_{j=1}^d \EE\big[\Psi_j[\sigma^{s_j}]\big].
\end{align*}
Note that 
\begin{align*}
&\EE\big[\Psi_j[\sigma^{s_j}]\big] = \Gamma_0^{-1} \big[\EE_{U_i, b_i}[U_i\ketbra{b_i}U_i^\dagger]\big]\\
&=\Gamma_0^{-1} \big[\sum_{U\in \set{ I, H, S^\dagger H}}\sum_{b\in \set{0,1}} \frac{1}{3}\matrixelement{b}{U^{\dagger}\sigma^{s_j} U}{b}~ U\ketbra{b}U^\dagger]\big]\\
&=\Gamma_0^{-1}[\Gamma[\sigma^{s_j}]]= \sigma^{s_j}. 
\end{align*}
As a result of this equation, $$\EE[\hat{\rho}_i] = \sum_{\bfs} \alpha_s  \tensor_{j=1}^d \sigma^{s_j} = \rho_i.$$ Hence the proof is complete.
\end{proof}
It follows from large deviation analysis that we get the same error bound as in \eqref{eq:Pauli estimation error} with the circuit in Figure \ref{fig:Shadow dqubit}. 
 Therefore, we obtain a scalable estimation circuit consisting of independent single-qubit quantum operations, resulting in a $O(d)$ gate complexity. With that in mind, we summarize our design and present Algorithm \ref{alg: QLD juntas}. It remains to complete the proof of Theorem \ref{thm: Kjunta sample complexity}.
    \begin{algorithm}[t]
\caption{Algorithm for  $k$-qubit Circuits}
\label{alg: QLD juntas}
\DontPrintSemicolon
\KwIn{$k\leq d$, and $n$ samples $(\rho_i, y_i)_{i=1}^n$.}
\KwOut{Predictor $\hat{\mathcal{M}}$}
 \SetKwProg{Fn}{}{:}{} 
 \SetKwFunction{EST}{PauliEstimation}
 \SetKwFunction{main}{LearningAlgorithm}
\Fn{\main}{ 
\Fn{\EST}{

\For{$i=1:n$}{
Choose $d$ unitary $U_j$ randomly from $\set{I,H, S^\dagger H}$.\;
Apply the circuit in Figure \ref{fig:Shadow dqubit} with the selected unitary operators.\;
}
\For{$\bfs \in \set{0,1,2, 3}^d$ with $|\supp(\bfs)|\leq k$}{
Compute $\qfsest$ using \eqref{eq:pauli estimate }.\;
}
}

\For{$\mathcal{J}\subset [d]$ with $|\mathcal{J}|=k$}{
Compute $\qfYestJ=\sum_{\bfs: \supp(\bfs) \subseteq \mathcal{J}} \qfsest \qps$.\;
Find $\hat{\mathcal{J}}$ that maximizes $\norm\big{\qfYestJ}_{1}$.\; 
}
Construct $\PiJhat$ as in \eqref{eq:Pi k} with $\mathcal{J}=\hat{\mathcal{J}}$.\;
 \KwRet  POVM $\hat{\mathcal{M}}:=\{\hat{\Pi}^{\hat{J}}, I-\hat{\Pi}^{\hat{J}}\}$. 
}
\end{algorithm} 
\paragraph{Proof of Theorem \ref{thm: Kjunta sample complexity}:} We show that $\hat{\mathcal{M}}$, the output of Algorithm  \ref{alg: QLD juntas} achieves the optimal loss $\opt_k$.
We use Theorem \ref{thm:loss M_k} followed by a Parseval-type identity. From \eqref{eq:Fourier trace}, it follows that $\norm{A}_2^2=\tr{A^\dagger A} = 2^d\sum_{\bfs} |a_\bfs|^2$.  Then, 
\begin{align*}
\Loss(\hat{\mathcal{M}}_{\mathcal{J}^*})&\leq \opt_k+4 ~2^d \sqrt{\sum_{\substack{\bfs: \supp(\bfs)\subseteq \mathcal{J}^*}}\hspace{-10pt}(\qfs-\qfsest)^2}.
\end{align*}
Hence, with Theorem  \ref{thm:Pauli Estimation Shadow}, we have that
\begin{align*}
\Loss(\mathcal{M}_{\clJ^*}) & \leq \Popt + 4 \sqrt{\sum_{\substack{\bfs: \supp(\bfs)\subseteq \mathcal{J}^*}}\hspace{-5pt}\order{{\frac{1}{n}\log(\frac{K}{\delta})}}}\\\numberthis\label{eq:L M J star}
&=\Popt + \order{\sqrt{\frac{4^k}{n}\log(\frac{K}{\delta})}},
\end{align*}
where we used the fact that $|\mathcal{J}^*|=k$. 
Note that $\mathcal{J}^*$ is unknown as it is defined based on the true operator $\qYJ$. We need to show that $\hat{\clJ}$ in Algorithm \ref{alg: QLD juntas} is ``close" to $\clJ^*$. 
From Lemma \ref{lem: k junta opt}, it suffices to show that $\norm\big{{\qfYest}^{ \hat{\mathcal{J}}}}_{1}$ is close to $\norm\big{{\qfY}^{ {\mathcal{J}^*}}}_{1}$ which gives $\opt_k$. Since $\hat{\clJ}$ maximizes $\norm\big{\qfYestJ}_1$, then $\norm\big{{\qfYest}^{\hat{\mathcal{J}}}}_{1} \geq \norm\big{{\qfYest}^{{\mathcal{J}}^*}}_{1}$. From the triangle inequality and the relation $\norm{\cdot}_1\leq \sqrt{dim}\norm{\cdot}_2$, we have that $$\abs{\norm\big{{\qfYest}^{{\mathcal{J}}^*}}_{1} -\norm\big{{\qfY}^{{\mathcal{J}}^*}}_{1} } \leq \sqrt{2^d}\norm\big{{\qfYest}^{{\mathcal{J}}^*} - {\qfY}^{{\mathcal{J}}^*}}_2\leq \epsilon_n,$$
where $\epsilon_n$ is the second term in \eqref{eq:L M J star}.  
The last inequality follows from Theorem \ref{thm:Pauli Estimation Shadow}.
Combining this inequality with Lemma \ref{lem:norm2 sign} and \ref{lem: k junta opt}, we obtain that
\begin{align*}
\Loss(\hat{\mathcal{M}})&\leq \Loss(\mathcal{M}_{\clJ^*}) +\epsilon_n =\Popt + 2 \epsilon_n, 
\end{align*}
where $\hat{\mathcal{M}}$ is the output of Algorithm \ref{alg: QLD juntas}. The proof is complete by replacing the expression for $\epsilon_n$ in given \eqref{eq:L M J star} and that of $K$ in \eqref{eq:K}.

 \begin{figure}[h!]
\centering
 \includegraphics[scale=0.2]{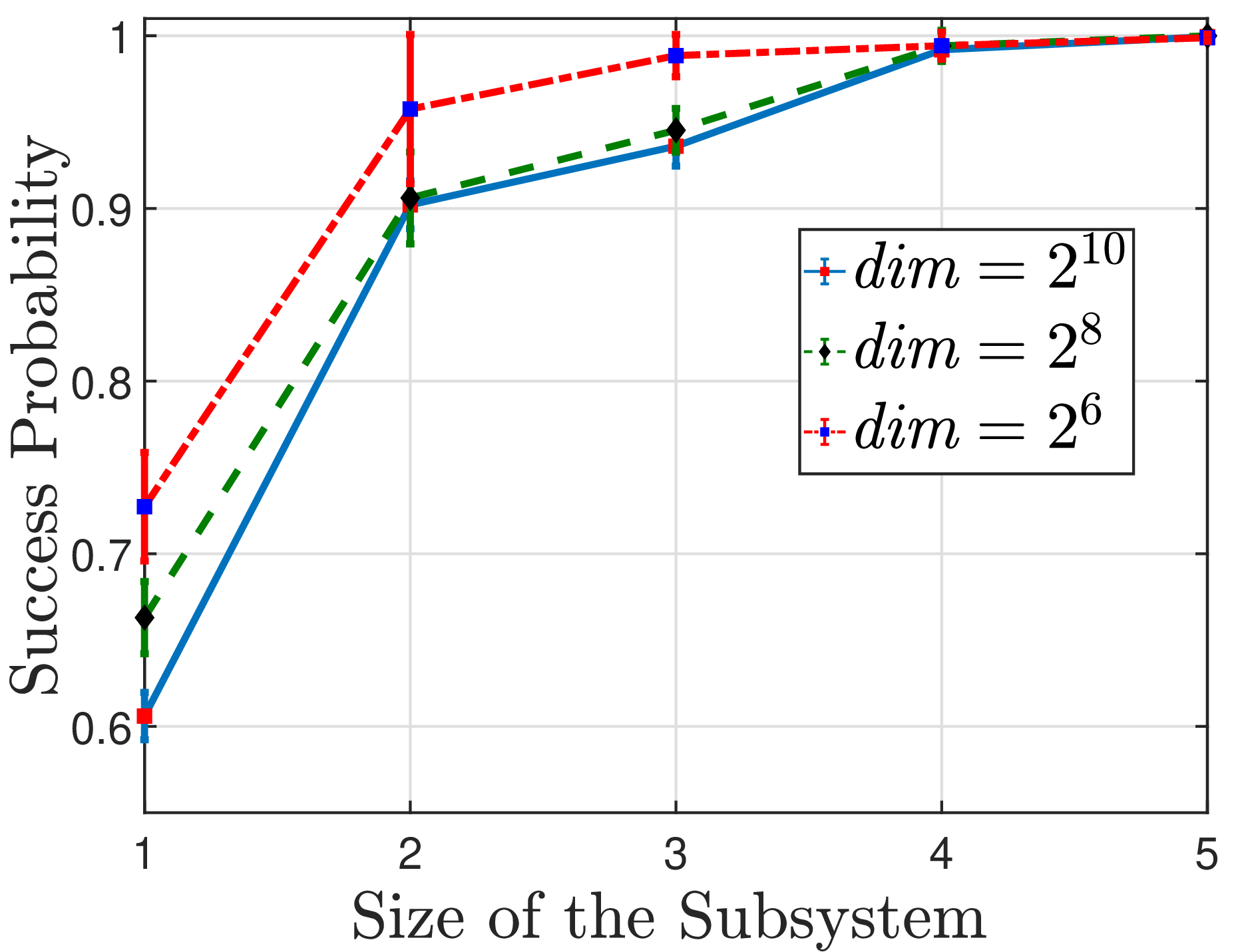} 
   \caption{Success probability vs. the size of the sub-system ($k$) for classifying maximally entangled and separable $d$-qubit states with different values of $d$.}
   \label{fig:junta}
 \end{figure}

 \section{Numerical Validation}\label{subse:numerical}
 We test Algorithm \ref{alg: QLD juntas} for classifying separable states from maximally entangled. For that, we generate a training data set by randomly generating $d$-qubit states. We generate two types of states: separable (with label $y=0$) and maximally entangled (with label $y=1$).  For that, we use \textit{RandomDensityMatrix} in \citep{qetlab} to generate a separable random density matrix based on a Haar measure.

 Figure \ref{fig:junta} shows the success probability versus the size of the sub-system ($k$) with different dimensions ($2^d$). Our results indicate that accessing only a small subsystem is sufficient to obtain a reasonable accuracy. For instance, a success probability of $0.95$ is possible using a $3$-qubit subsystem inside the original $10$-qubit system. 

\begin{figure}
\centering
    \includegraphics[scale =0.35]{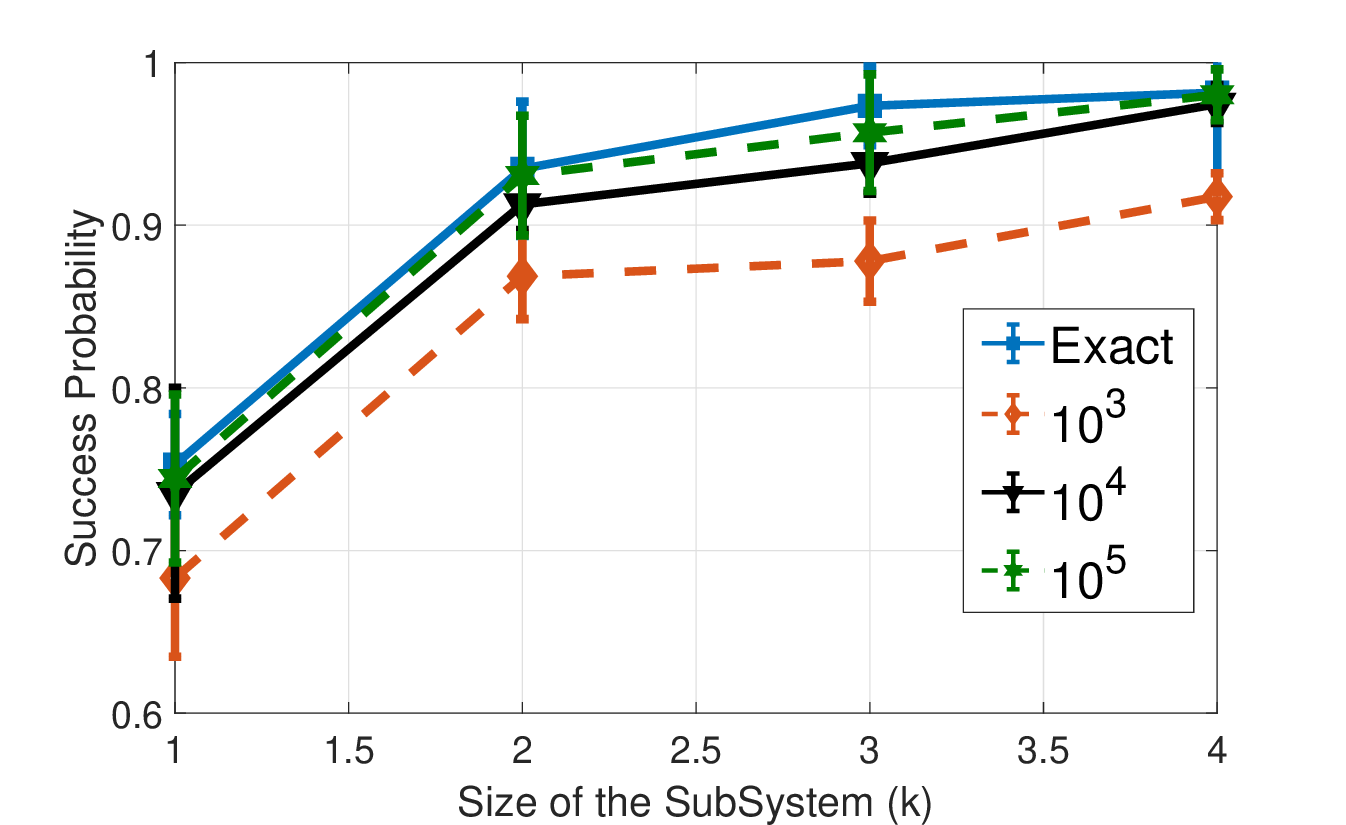}

  \caption{Success probability of Algorithm \ref{alg: QLD juntas} for various $k$ (subsystem size) and various samples $n=10^3, 10^4, 10^5$ compared to theoretical bound (exact). The experiment is averaged over $5$ runs with error bars showing the deviations.}
  \label{fig:algorithm sample}
\end{figure}

  Moreover, we tested the accuracy of Algorithm \ref{alg: QLD juntas} with different values of $k$ versus various sample sizes ($n = 10^3, 10^4,$ and $10^5$). The dataset is the same as in the previous experiment but with a fixed dimension  $dim=2^6$.  Figure \ref{fig:algorithm sample} demonstrates  success probability as a function of $k$ for various sample sizes. As observed, with more samples, the success probability converges to the theoretical values with exact computations.

\section*{Discussion and Future Directions}
We prove that the quantum sample complexity of $k$-qubit quantum operations  is $O(\frac{k4^k}{\epsilon^2}\log d)$ which grows logarithmically with the number of qubits and is comparable with the classical sample complexity of $k$-juntas. This is a surprising result due to the no-cloning principle, measurement incompatibility, richness of $k$-qubit operations, and the fact that QPAC is a stronger condition than classical PAC. We propose a  quantum algorithm that provably  QPAC learns $k$-qubit operations. We develop a new connection to Pauli decomposition with a new estimation method with a scalable circuit.

Our results indicate that shallow-width quantum circuits are learnable with a sample complexity growing logarithmically with the number of qubits ($d$). In future work, one can study learning of shallow-depth quantum circuits and compare the quantum sample complexity of such circuits with constant-depth classical circuits. Whether quantum sample complexity is comparable with the classical one is an important direction to pursue.  

 


\section*{Acknowledgments}
This work was partially supported by the NSF Center for Science of Information (CSoI) Grant
CCF-0939370, and also by NSF Grants CCF-2006440, CCF-2007238, CCF-2211423, and Google Research Award.

\bibliography{main}

\begin{thebibliography}{54}
\providecommand{\natexlab}[1]{#1}
\providecommand{\url}[1]{\texttt{#1}}
\expandafter\ifx\csname urlstyle\endcsname\relax
  \providecommand{\doi}[1]{doi: #1}\else
  \providecommand{\doi}{doi: \begingroup \urlstyle{rm}\Url}\fi

\bibitem[Aaronson(2007)]{Aaronson2007}
S.~Aaronson.
\newblock The learnability of quantum states.
\newblock \emph{Proceedings of the Royal Society A: Mathematical, Physical and
  Engineering Sciences}, 463\penalty0 (2088):\penalty0 3089--3114, sep 2007.
\newblock \doi{10.1098/rspa.2007.0113}.

\bibitem[Aaronson(2018)]{Aaronson2018}
S.~Aaronson.
\newblock Shadow tomography of quantum states.
\newblock In \emph{Proceedings of the 50th Annual ACM SIGACT Symposium on
  Theory of Computing}, STOC 2018, page 325–338, New York, NY, USA, 2018.
  Association for Computing Machinery.
\newblock ISBN 9781450355599.
\newblock \doi{10.1145/3188745.3188802}.

\bibitem[Arunachalam and de~Wolf(2017)]{Arunachalam2017}
S.~Arunachalam and R.~de~Wolf.
\newblock A survey of quantum learning theory.
\newblock \emph{arXiv:1701.06806}, 2017.

\bibitem[Arunachalam and De~Wolf(2018)]{Arunachalam2018}
S.~Arunachalam and R.~De~Wolf.
\newblock Optimal quantum sample complexity of learning algorithms.
\newblock \emph{J. Mach. Learn. Res.}, 19\penalty0 (1):\penalty0 2879–2878,
  Jan. 2018.
\newblock ISSN 1532-4435.

\bibitem[Audenaert et~al.(2008)Audenaert, Nussbaum, Szko{\l}a, and
  Verstraete]{Audenaert2008}
K.~M.~R. Audenaert, M.~Nussbaum, A.~Szko{\l}a, and F.~Verstraete.
\newblock Asymptotic error rates in quantum hypothesis testing.
\newblock \emph{Communications in Mathematical Physics}, 279\penalty0
  (1):\penalty0 251--283, feb 2008.
\newblock \doi{10.1007/s00220-008-0417-5}.

\bibitem[Badescu et~al.(2019)Badescu, O'Donnell, and Wright]{Badescu2019}
C.~Badescu, R.~O'Donnell, and J.~Wright.
\newblock Quantum state certification.
\newblock In \emph{Proceedings of the 51st Annual {ACM} {SIGACT} Symposium on
  Theory of Computing}. {ACM}, jun 2019.
\newblock \doi{10.1145/3313276.3316344}.

\bibitem[Barnett and Croke(2009)]{Barnett2009}
S.~M. Barnett and S.~Croke.
\newblock Quantum state discrimination.
\newblock \emph{Advances in Optics and Photonics}, 1\penalty0 (2):\penalty0
  238, feb 2009.
\newblock \doi{10.1364/aop.1.000238}.

\bibitem[Bauer et~al.(2020)Bauer, Bravyi, Motta, and Chan]{Bauer2020}
B.~Bauer, S.~Bravyi, M.~Motta, and G.~K.-L. Chan.
\newblock Quantum algorithms for quantum chemistry and quantum materials
  science.
\newblock \emph{Chemical Reviews}, 120\penalty0 (22):\penalty0 12685--12717,
  oct 2020.
\newblock \doi{10.1021/acs.chemrev.9b00829}.

\bibitem[Bernstein and Vazirani(1997)]{Bernstein1997}
E.~Bernstein and U.~Vazirani.
\newblock Quantum complexity theory.
\newblock \emph{{SIAM} Journal on Computing}, 26\penalty0 (5):\penalty0
  1411--1473, oct 1997.
\newblock \doi{10.1137/s0097539796300921}.

\bibitem[Biamonte et~al.(2017)Biamonte, Wittek, Pancotti, Rebentrost, Wiebe,
  and Lloyd]{Biamonte2017}
J.~Biamonte, P.~Wittek, N.~Pancotti, P.~Rebentrost, N.~Wiebe, and S.~Lloyd.
\newblock Quantum machine learning.
\newblock 549\penalty0 (7671):\penalty0 195--202, sep 2017.
\newblock \doi{10.1038/nature23474}.

\bibitem[Broecker et~al.(2017)Broecker, Carrasquilla, Melko, and
  Trebst]{Broecker2017}
P.~Broecker, J.~Carrasquilla, R.~G. Melko, and S.~Trebst.
\newblock Machine learning quantum phases of matter beyond the fermion sign
  problem.
\newblock 7\penalty0 (1), aug 2017.
\newblock \doi{10.1038/s41598-017-09098-0}.

\bibitem[Broughton et~al.(2020)Broughton, Verdon, McCourt, Martinez, Yoo,
  Isakov, Massey, Halavati, Niu, Zlokapa, Peters, Lockwood, Skolik, Jerbi,
  Dunjko, Leib, Streif, Dollen, Chen, Cao, Wiersema, Huang, McClean, Babbush,
  Boixo, Bacon, Ho, Neven, and Mohseni]{Broughton2020}
M.~Broughton, G.~Verdon, T.~McCourt, A.~J. Martinez, J.~H. Yoo, S.~V. Isakov,
  P.~Massey, R.~Halavati, M.~Y. Niu, A.~Zlokapa, E.~Peters, O.~Lockwood,
  A.~Skolik, S.~Jerbi, V.~Dunjko, M.~Leib, M.~Streif, D.~V. Dollen, H.~Chen,
  S.~Cao, R.~Wiersema, H.-Y. Huang, J.~R. McClean, R.~Babbush, S.~Boixo,
  D.~Bacon, A.~K. Ho, H.~Neven, and M.~Mohseni.
\newblock Tensorflow quantum: A software framework for quantum machine
  learning.
\newblock \emph{arXiv:2003.02989}, Mar. 2020.

\bibitem[Bshouty and Jackson(1998)]{Bshouty1998}
N.~H. Bshouty and J.~C. Jackson.
\newblock Learning dnf over the uniform distribution using a quantum example
  oracle.
\newblock \emph{SIAM Journal on Computing}, 28\penalty0 (3):\penalty0
  1136--1153, 1998.

\bibitem[Bubeck et~al.(2020)Bubeck, Chen, and Li]{Bubeck2020}
S.~Bubeck, S.~Chen, and J.~Li.
\newblock Entanglement is necessary for optimal quantum property testing.
\newblock In \emph{2020 {IEEE} 61st Annual Symposium on Foundations of Computer
  Science ({FOCS})}. {IEEE}, nov 2020.
\newblock \doi{10.1109/focs46700.2020.00070}.

\bibitem[Cao et~al.(2019)Cao, Romero, Olson, Degroote, Johnson,
  Kieferov{\'{a}}, Kivlichan, Menke, Peropadre, Sawaya, Sim, Veis, and
  Aspuru-Guzik]{Cao2019}
Y.~Cao, J.~Romero, J.~P. Olson, M.~Degroote, P.~D. Johnson, M.~Kieferov{\'{a}},
  I.~D. Kivlichan, T.~Menke, B.~Peropadre, N.~P.~D. Sawaya, S.~Sim, L.~Veis,
  and A.~Aspuru-Guzik.
\newblock Quantum chemistry in the age of quantum computing.
\newblock \emph{Chemical Reviews}, 119\penalty0 (19):\penalty0 10856--10915,
  aug 2019.
\newblock \doi{10.1021/acs.chemrev.8b00803}.

\bibitem[Carleo and Troyer(2017)]{Carleo2017}
G.~Carleo and M.~Troyer.
\newblock Solving the quantum many-body problem with artificial neural
  networks.
\newblock 355\penalty0 (6325):\penalty0 602--606, feb 2017.
\newblock \doi{10.1126/science.aag2302}.

\bibitem[Carrasquilla and Melko(2017)]{Carrasquilla2017}
J.~Carrasquilla and R.~G. Melko.
\newblock Machine learning phases of matter.
\newblock 13\penalty0 (5):\penalty0 431--434, feb 2017.
\newblock \doi{10.1038/nphys4035}.

\bibitem[Chen et~al.(2021)Chen, Ren, Lin, and Lu]{Chen2021a}
C.~Chen, C.~Ren, H.~Lin, and H.~Lu.
\newblock Entanglement structure detection via machine learning.
\newblock \emph{Quantum Science and Technology}, 2021.

\bibitem[Cheng et~al.(2015)Cheng, Hsieh, and Yeh]{Cheng2015}
H.-C. Cheng, M.-H. Hsieh, and P.-C. Yeh.
\newblock The learnability of unknown quantum measurements.
\newblock \emph{QIC, Vol. 16, No. 7-8, 0615-0656 (2016)}, Jan. 2015.

\bibitem[Crawford et~al.(2020)Crawford, van Straaten, Wang, Parks, Campbell,
  and Brierley]{Crawford2019}
O.~Crawford, B.~van Straaten, D.~Wang, T.~Parks, E.~Campbell, and S.~Brierley.
\newblock Efficient quantum measurement of pauli operators in the presence of
  finite sampling error.
\newblock \emph{arXiv:1908.06942}, 2020.

\bibitem[Deng et~al.(2017)Deng, Li, and Sarma]{Deng2017}
D.-L. Deng, X.~Li, and S.~D. Sarma.
\newblock Quantum entanglement in neural network states.
\newblock \emph{Physical Review X}, 7\penalty0 (2):\penalty0 021021, 2017.

\bibitem[Gambs(2008)]{Gambs2008}
S.~Gambs.
\newblock Quantum classification.
\newblock \emph{0809.0444 [quant-ph]}, Sept. 2008.

\bibitem[Giovannetti et~al.(2008)Giovannetti, Lloyd, and
  Maccone]{Giovannetti2008}
V.~Giovannetti, S.~Lloyd, and L.~Maccone.
\newblock Quantum random access memory.
\newblock \emph{Physical Review Letters}, 100\penalty0 (16), apr 2008.
\newblock \doi{10.1103/physrevlett.100.160501}.

\bibitem[Guta and Kotlowski(2010)]{Guta2010}
M.~Guta and W.~Kotlowski.
\newblock Quantum learning: asymptotically optimal classification of qubit
  states.
\newblock \emph{New Journal of Physics}, 12\penalty0 (12):\penalty0 123032, dec
  2010.
\newblock \doi{10.1088/1367-2630/12/12/123032}.

\bibitem[Haah et~al.(2016)Haah, Harrow, Ji, Wu, and Yu]{Haah2016}
J.~Haah, A.~W. Harrow, Z.~Ji, X.~Wu, and N.~Yu.
\newblock Sample-optimal tomography of quantum states.
\newblock In \emph{Proceedings of the forty-eighth annual {ACM} symposium on
  Theory of Computing}. {ACM}, jun 2016.
\newblock \doi{10.1145/2897518.2897585}.

\bibitem[Heidari et~al.(2021)Heidari, Padakandla, and
  Szpankowski]{HeidariQuantum2021}
M.~Heidari, A.~Padakandla, and W.~Szpankowski.
\newblock A theoretical framework for learning from quantum data.
\newblock In \emph{2021 {IEEE} International Symposium on Information Theory
  ({ISIT})}. {IEEE}, jul 2021.
\newblock \doi{10.1109/isit45174.2021.9517721}.

\bibitem[Heidari et~al.(2022)Heidari, Grama, and Szpankowski]{HeidariAAAI2022}
M.~Heidari, A.~Grama, and W.~Szpankowski.
\newblock Toward physically realizable quantum neural networks.
\newblock \emph{Proceedings of the {AAAI} Conference on Artificial
  Intelligence}, 36\penalty0 (6):\penalty0 6902--6909, jun 2022.
\newblock \doi{10.1609/aaai.v36i6.20647}.

\bibitem[Hempel et~al.(2018)Hempel, Maier, Romero, McClean, Monz, Shen,
  Jurcevic, Lanyon, Love, Babbush, Aspuru-Guzik, Blatt, and Roos]{Hempel2018}
C.~Hempel, C.~Maier, J.~Romero, J.~McClean, T.~Monz, H.~Shen, P.~Jurcevic,
  B.~P. Lanyon, P.~Love, R.~Babbush, A.~Aspuru-Guzik, R.~Blatt, and C.~F. Roos.
\newblock Quantum chemistry calculations on a trapped-ion quantum simulator.
\newblock \emph{Physical Review X}, 8\penalty0 (3):\penalty0 031022, jul 2018.
\newblock \doi{10.1103/physrevx.8.031022}.

\bibitem[Hiesmayr(2021)]{Hiesmayr2021}
B.~C. Hiesmayr.
\newblock Free versus bound entanglement, a {NP}-hard problem tackled by
  machine learning.
\newblock \emph{Scientific Reports}, 11\penalty0 (1), oct 2021.
\newblock \doi{10.1038/s41598-021-98523-6}.

\bibitem[Huang et~al.(2020)Huang, Kueng, and Preskill]{Huang2020}
H.-Y. Huang, R.~Kueng, and J.~Preskill.
\newblock Predicting many properties of a quantum system from very few
  measurements.
\newblock \emph{Nature Physics 16, 1050--1057 (2020)}, Feb. 2020.
\newblock \doi{10.1038/s41567-020-0932-7}.

\bibitem[Johnston et~al.(2016)Johnston, Cosentino, and Russo]{qetlab}
N.~Johnston, A.~Cosentino, and V.~Russo.
\newblock Qetlab: Qetlab v0.9, 2016.

\bibitem[Kanade et~al.(2019)Kanade, Rocchetto, and Severini]{Kanade2018}
V.~Kanade, A.~Rocchetto, and S.~Severini.
\newblock Learning dnfs under product distributions via $\mu$-biased quantum
  fourier sampling.
\newblock \emph{arXiv:1802.05690v3}, 2019.

\bibitem[Kassal et~al.(2011)Kassal, Whitfield, Perdomo-Ortiz, Yung, and
  Aspuru-Guzik]{Kassal2011}
I.~Kassal, J.~D. Whitfield, A.~Perdomo-Ortiz, M.-H. Yung, and A.~Aspuru-Guzik.
\newblock Simulating chemistry using quantum computers.
\newblock \emph{Annual Review of Physical Chemistry}, 62\penalty0 (1):\penalty0
  185--207, may 2011.
\newblock \doi{10.1146/annurev-physchem-032210-103512}.

\bibitem[Kearns et~al.(1994)Kearns, Schapire, and Sellie]{Kearns1994}
M.~J. Kearns, R.~E. Schapire, and L.~M. Sellie.
\newblock Toward efficient agnostic learning.
\newblock \emph{Machine Learning}, 17\penalty0 (2-3):\penalty0 115--141, 1994.
\newblock \doi{10.1007/bf00993468}.

\bibitem[Lloyd et~al.(2013)Lloyd, Mohseni, and Rebentrost]{Lloyd2013}
S.~Lloyd, M.~Mohseni, and P.~Rebentrost.
\newblock Quantum algorithms for supervised and unsupervised machine learning.
\newblock \emph{arXiv:1307.0411}, 2013.

\bibitem[Lloyd et~al.(2014)Lloyd, Mohseni, and Rebentrost]{Lloyd2014}
S.~Lloyd, M.~Mohseni, and P.~Rebentrost.
\newblock Quantum principal component analysis.
\newblock 10\penalty0 (9):\penalty0 631--633, jul 2014.
\newblock \doi{10.1038/nphys3029}.

\bibitem[Lu et~al.(2018)Lu, Huang, Li, Li, Chen, Lu, Ji, Shen, Zhou, and
  Zeng]{Lu2018}
S.~Lu, S.~Huang, K.~Li, J.~Li, J.~Chen, D.~Lu, Z.~Ji, Y.~Shen, D.~Zhou, and
  B.~Zeng.
\newblock Separability-entanglement classifier via machine learning.
\newblock \emph{Physical Review A}, 98\penalty0 (1):\penalty0 012315, 2018.

\bibitem[Ma and Yung(2018)]{Ma2018}
Y.-C. Ma and M.-H. Yung.
\newblock Transforming bell's inequalities into state classifiers with machine
  learning.
\newblock \emph{npj Quantum Information}, 4\penalty0 (1), jul 2018.
\newblock \doi{10.1038/s41534-018-0081-3}.

\bibitem[Massoli et~al.(2021)Massoli, Vadicamo, Amato, and
  Falchi]{massoli2021leap}
F.~V. Massoli, L.~Vadicamo, G.~Amato, and F.~Falchi.
\newblock A leap among entanglement and neural networks: A quantum survey.
\newblock \emph{arXiv:2107.03313}, July 2021.

\bibitem[McArdle et~al.(2020)McArdle, Endo, Aspuru-Guzik, Benjamin, and
  Yuan]{McArdle2020}
S.~McArdle, S.~Endo, A.~Aspuru-Guzik, S.~C. Benjamin, and X.~Yuan.
\newblock Quantum computational chemistry.
\newblock \emph{Reviews of Modern Physics}, 92\penalty0 (1):\penalty0 015003,
  mar 2020.
\newblock \doi{10.1103/revmodphys.92.015003}.

\bibitem[Montanaro and de~Wolf(2016)]{Montanaro2016}
A.~Montanaro and R.~de~Wolf.
\newblock A survey of quantum property testing.
\newblock \emph{Theory of Computing}, 1\penalty0 (1):\penalty0 1--81, 2016.
\newblock \doi{10.4086/toc.gs.2016.007}.

\bibitem[Montanaro and Osborne(2010)]{Montanaro2008}
A.~Montanaro and T.~J. Osborne.
\newblock Quantum boolean functions.
\newblock \emph{arXiv:0810.2435}, 2010.

\bibitem[Mossel et~al.(2003)Mossel, O'Donnell, and Servedio]{Mossel_ODonnell}
E.~Mossel, R.~O'Donnell, and R.~P. Servedio.
\newblock Learning juntas.
\newblock In \emph{Proc. ACM Symp. on Theory of Computing}, pages 206--212,
  2003.

\bibitem[Mossel et~al.(2004)Mossel, O'Donnell, and
  Servedio]{mossel2004learning}
E.~Mossel, R.~O'Donnell, and R.~A. Servedio.
\newblock Learning functions of $k$ relevant variables.
\newblock \emph{J. Comput. Syst. Sci}, 69\penalty0 (3):\penalty0 421--434,
  2004.

\bibitem[O'Donnell and Wright(2016)]{ODonnell2016}
R.~O'Donnell and J.~Wright.
\newblock Efficient quantum tomography.
\newblock In \emph{Proceedings of the forty-eighth annual {ACM} symposium on
  Theory of Computing}. {ACM}, jun 2016.
\newblock \doi{10.1145/2897518.2897544}.

\bibitem[O'Donnell and Wright(2017)]{ODonnell2017}
R.~O'Donnell and J.~Wright.
\newblock Efficient quantum tomography {II}.
\newblock In \emph{Proceedings of the 49th Annual {ACM} {SIGACT} Symposium on
  Theory of Computing}. {ACM}, jun 2017.
\newblock \doi{10.1145/3055399.3055454}.

\bibitem[Park et~al.(2019)Park, Petruccione, and Rhee]{Park2019}
D.~K. Park, F.~Petruccione, and J.-K.~K. Rhee.
\newblock Circuit-based quantum random access memory for classical data.
\newblock \emph{Scientific Reports}, 9\penalty0 (1), mar 2019.
\newblock \doi{10.1038/s41598-019-40439-3}.

\bibitem[Peruzzo et~al.(2014)Peruzzo, McClean, Shadbolt, Yung, Zhou, Love,
  Aspuru-Guzik, and O'Brien]{Peruzzo2014}
A.~Peruzzo, J.~McClean, P.~Shadbolt, M.-H. Yung, X.-Q. Zhou, P.~J. Love,
  A.~Aspuru-Guzik, and J.~L. O'Brien.
\newblock A variational eigenvalue solver on a photonic quantum processor.
\newblock \emph{Nature Communications}, 5\penalty0 (1), jul 2014.
\newblock \doi{10.1038/ncomms5213}.

\bibitem[Rebentrost et~al.(2014)Rebentrost, Mohseni, and Lloyd]{Rebentrost2014}
P.~Rebentrost, M.~Mohseni, and S.~Lloyd.
\newblock Quantum support vector machine for big data classification.
\newblock \emph{Physical Review Letters}, 113\penalty0 (13), sep 2014.
\newblock \doi{10.1103/physrevlett.113.130503}.

\bibitem[Schuld et~al.(2014)Schuld, Sinayskiy, and Petruccione]{Schuld_2014}
M.~Schuld, I.~Sinayskiy, and F.~Petruccione.
\newblock The quest for a quantum neural network.
\newblock \emph{Quantum Information Processing}, 13\penalty0 (11):\penalty0
  2567–2586, Aug 2014.
\newblock ISSN 1573-1332.
\newblock \doi{10.1007/s11128-014-0809-8}.

\bibitem[Schuld et~al.(2020)Schuld, Bocharov, Svore, and Wiebe]{Schuld_2020}
M.~Schuld, A.~Bocharov, K.~M. Svore, and N.~Wiebe.
\newblock Circuit-centric quantum classifiers.
\newblock \emph{Physical Review A}, 101\penalty0 (3):\penalty0 032308, mar
  2020.
\newblock \doi{10.1103/physreva.101.032308}.

\bibitem[Servedio and Gortler(2004)]{Servedio2004}
R.~A. Servedio and S.~J. Gortler.
\newblock Equivalences and separations between quantum and classical
  learnability.
\newblock \emph{SIAM J. Comput.}, 33\penalty0 (5):\penalty0 1067–1092, May
  2004.
\newblock ISSN 0097-5397.
\newblock \doi{10.1137/S0097539704412910}.
\newblock URL \url{https://doi.org/10.1137/S0097539704412910}.

\bibitem[Shalev-Shwartz and Ben-David(2014)]{ShalevShwartz2014}
S.~Shalev-Shwartz and S.~Ben-David.
\newblock \emph{Understanding Machine Learning: From Theory to Algorithms}.
\newblock Cambridge University Press, New York, NY, USA, 2014.
\newblock ISBN 1107057132, 9781107057135.

\bibitem[Valiant(1984)]{Valiant1984}
L.~G. Valiant.
\newblock A theory of the learnable.
\newblock \emph{Communications of the {ACM}}, 27\penalty0 (11):\penalty0
  1134--1142, nov 1984.
\newblock \doi{10.1145/1968.1972}.

\end{thebibliography}

\newpage
\onecolumn
\appendix
\section{Proof of The Technical Lemmas}\label{sec:proofs}
\subsection{Proof of Lemma \ref{lem: k junta opt}}\label{proof:lem: k junta opt}
\begin{proof}
 We start with proving a lower bound on $\opt_k$. Fix a $k$ element coordinate subset $\mathcal{I}\subset [d]$ and consider a $k$-qubit measurement that depends only on coordinates $\mathcal{I}$. From Lemma \ref{thm:mislabeling prob}, the loss of $\mathcal{M}$ equals to 
 \begin{align*}
\Loss(\mathcal{M}) = \frac{1}{2}-2^{d-1}\sum_{\bfs} \qfs \qgs. 
 \end{align*}
  From Definition \ref{def:k-qubit measurement}, as $\mathcal{M}$ depends only on coordinate $\mathcal{I}$, then one can show that $\qgs=0$ for all $\bfs$ with $\supp(\bfs)\nsubseteq \mathcal{I}$. Therefore, 
\begin{align*}
\Loss(\mathcal{M}) = \frac{1}{2}-2^{d-1}\sum_{\bfs: \supp(\bfs)\subseteq \mathcal{I}} \qfs \qgs.
  \end{align*}  
 Define the following operator: 
\begin{align}\label{eq:Loss M_I}
\qfY^{\mathcal{I}} := \sum_{\bfs: \supp(\bfs)\subseteq \mathcal{I}}  \qfs \qps.
 \end{align} 
Therefore, from \eqref{eq:Fourier trace}, we have that 
\begin{align*}
\Loss(\mathcal{M}) &= \frac{1}{2}-\frac{1}{2} \tr{\qfY^{\mathcal{I}}  G} \geq \frac{1}{2}-\frac{1}{2} \tr{\abs{\qfY^{\mathcal{I}}}  \abs{G}}=\frac{1}{2}-\frac{1}{2} \tr{\abs{\qfY^{\mathcal{I}}}},
  \end{align*} 
  where $\abs{A} = \sqrt{A^\dagger A}$ and  the first inequality follows as $\tr{A}\leq \tr{\abs{A}}$. The second equality follows as the eigenvalues of $G$ belong to $\pmm$, implying that $\abs{G}=\id$. Therefore,  we obtain that 
 
 \begin{align*}
 \Loss(\mathcal{M})\geq \frac{1}{2}-\frac{1}{2} \norm{\qfY^{\mathcal{I}}}_{1}\geq  \frac{1}{2}-\frac{1}{2} \max_{\mathcal{J}\subset [d]: |\mathcal{J}|= k}\norm{\qYJ}_{1},
 \end{align*}
 where the last inequality holds by minimizing the lower bound over all $k$-element coordinates $\mathcal{J}$. Note that the above bound holds for all  $\mathcal{M}$ depending on any $k$-element coordinate subset $\mathcal{I}$. Thus, we obtain the lower bound on $\opt_k$:
   \begin{align}\label{eq: opt k lb}
   \opt_k\geq \frac{1}{2}-\frac{1}{2} \max_{\mathcal{J}\subset [d]: |\mathcal{J}|= k}\norm{\qYJ}_{1}.
   \end{align}
 Next, we establish the achievability of the lower bound. Again fix a $k$-element subset $\mathcal{J}\subset [d]$ and let $G_{M_{\mathcal{J}}}=\sign[\qYJ]$. Note that we can consider a valid measurement $\mathcal{M}_{\mathcal{J}}$ corresponding to $G_{M_{\mathcal{J}}}$. Moreover, $G_{M_{\mathcal{J}}}$ is a $k$-junta operator depending only on the coordinates $\mathcal{J}$. Therefore, its Fourier coefficients $\qgs$ are zero for any $\bfs$ with $\supp(\bfs)\nsubseteq \mathcal{J}$.  As a result, from \eqref{eq:Loss M_I} 
 \begin{align*}
 \Loss(\mathcal{M}_{\mathcal{J}})&=\frac{1}{2}-\frac{1}{2}\tr{G_{M_{\mathcal{J}}} \qYJ}\\
 &=\frac{1}{2}-\frac{1}{2}\tr{\sign[\qfY^{\mathcal{J}}] \qYJ}\\
 &=\frac{1}{2}-\frac{1}{2} \norm{\qYJ}_{1},
 \end{align*}
 where the last equality follows from the identity $\norm{A}_{1}=\tr{A \sign[A]}$ that holds for any Hermitian and bounded operator $A$. With the above inequality, optimizing over $\mathcal{J}$ gives
   \begin{align}\label{eq:opt k up}
   \min_{\mathcal{J}\subset [d]: |\mathcal{J}|= k}  \Loss(\mathcal{M}_{\mathcal{J}}) = \frac{1}{2}-\frac{1}{2} \max_{\mathcal{J}\subset [d]: |\mathcal{J}|= k}\norm{\qYJ}_{1}.
   \end{align}
   Note that $ \opt_k$ is smaller than the left-hand side of \eqref{eq:opt k up}, as it has an additional minimization over the choice of the measurements: $$\opt_k=\min_{\mathcal{J}\subset [d]: |\mathcal{J}|= k} \inf_{\mathcal{M}_{\mathcal{J}}}\Loss(\mathcal{M}_{\mathcal{J}}).$$
 Therefore, \eqref{eq:opt k up} is an upper bound for $\opt(k)$. As this upper bound matches with the lower bound in \eqref{eq: opt k lb}, then we obtain the equality in \eqref{eq:opt k up}. This proves the expression for $\opt(k)$ and that $\mathcal{M}_{\mathcal{J}^*}$ is the best $k$-junta measurement.

\end{proof}
\subsection{Proof of Lemma \ref{lem:norm2 sign}}\label{app:proof norm2 sign}
Let  $\GJ = M_1-M_0= I-2\PiJhat$. It is not difficult to check that $\GJ=\sign[\qfYestJ]$. From Lemma \ref{thm:mislabeling prob} in the main text, the loss of $M_{\mathcal{J}}$ can be written as $\Loss(\mathcal{M}_{\mathcal{J}}) = \frac{1}{2}-2^{d-1}\sum_\bfs \qfs\qgs$, where $\qfs$ and $\qgs$ are the Pauli coefficients of $\qfY$ and $\GJ$. Note that $\GJ$ depends only on the coordinates of $\mathcal{J}$. More precisely, $\GJ = (I^{\mathcal{J}}- 2\PiJhat)\tensor I^{\mathcal{J}^c}$, where $I^{\mathcal{J}}$  and $I^{\mathcal{J}^c}$ are the identity operators on the corresponding systems. Hence, the Pauli coefficients of $\qgs$ of $\GJ$ are zero outside of $\mathcal{J}$. Therefore, 
\begin{align*}\nonumber
\Loss(\mathcal{M}_{\mathcal{J}}) &= \frac{1}{2}-2^{d-1}\sum_{\bfs: \supp(\bfs)\subseteq \mathcal{J}} \qfs \qgs
\end{align*}
Define the $2$-norm of an operator $A$ as $\norm{A}_2 : = \sqrt{\tr{A^\dagger A}}$. Then, from \eqref{eq:Fourier trace}, $\norm{A}^2_2=2^d\sum_\bfs |a_\bfs|^2$. Moreover, for any pair of Hermitian operators $A, B$ we have the identity $\norm{A-B}_2^2=\norm{A}_2^2+\norm{B}_2^2-2\tr{A B}$. Therefore, from \eqref{eq:Fourier trace}, we obtain that 
\begin{align*}
\sum_\bfs a_\bfs b_\bfs = 2^{-d}\tr{A B}=2^{-d-1} \Big(\norm{A}_2^2 +\norm{B}_2^2 - \norm{A-B}_2^2\Big).
\end{align*}
Therefore, from the definition of $\qYJ$, we have that $$\sum_{\bfs: \supp(\bfs)\subseteq \mathcal{J}} \qfs \qgs = 2^{-d}\tr{\qYJ \GJ}.$$
As a result, the loss of $\mathcal{M}_\mathcal{J}$ can be written as
\begin{align}\nonumber
\Loss(\mathcal{M}_{\mathcal{J}}) &= \frac{1}{2}-\frac{1}{4}\Big(\norm{\GJ}_2^2+  \norm{\qYJ}_2^2-\norm{\qYJ-\GJ}_2^2  \Big)\\\label{eq:pe 2 norm stoch}
 &= \frac{1}{4}\big(2-2^d-\norm{\qYJ}_2^2+\norm{\qYJ-\GJ}_2^2\big),
\end{align}
where we used the fact that $\norm{\GJ}_2^2=2^d$ as the eigenvalues of $\GJ$ belong to $\pmm$.

Next, we bound the last $2$-norm quantity above. Recall that $\qfYestJ :=\sum_{\bfs: \supp(\bfs)\subseteq \mathcal{J}} \qfsest \qps$ is an approximation of $\qYJ$ using the estimated Pauli coefficients. By adding and subtracting $\qfYestJ$, we have that
\begin{align*}
 \norm{\qYJ-\GJ}_2^2  &\stackrel{(a)}{\leq} \Big(\norm{\qYJ-\qfYestJ}_2+\norm{\qfYestJ-\GJ}_2\Big)^2,\\\numberthis \label{eq:mismatch_2norm}
 &= \norm{\qYJ-\qfYestJ}^2_2+\norm{\qfYestJ-\GJ}^2_2+2\norm{\qYJ-\qfYestJ}_2\norm{\qfYestJ-\GJ}_2 ,
\end{align*}
where $(a)$  follows from the Minkowski's Inequality inequality for $2$-norm.  Note that $\GJ=\sign[\qfYestJ]$. Moreover, note that for any function $h$ the identity  $|h-\sign[h]|= |1-|h||$ holds. Therefore,
\begin{align}\nonumber
\norm{\qfYestJ-\GJ}^2_2&=\norm{\id-|\qfYestJ|}_2^2=\norm{\id}_2^2+\norm{\qfYestJ}_2^2-2\norm{\qfYestJ}_1\\\label{eq:fbar-sign}
&= 2^d+\norm{\qfYestJ}_2^2-2\norm{\qfYestJ}_1.
\end{align}
From this relation and equations \eqref{eq:pe 2 norm stoch},  \eqref{eq:mismatch_2norm}, we obtain the following upper bound
\begin{align*}
4\Loss(\mathcal{M}_{\mathcal{J}})&\leq 2-2^d-\norm{\qYJ}_2^2+\norm{\qYJ-\qfYestJ}^2_2+\norm{\qfYestJ-\GJ}^2_2+2\norm{\qYJ-\qfYestJ}_2\norm{\qfYestJ-\GJ}_2 \\
&= 2-2^d-\norm{\qYJ}_2^2+\norm{\qYJ-\qfYestJ}^2_2+2^d+\norm{\qfYestJ}_2^2-2\norm{\qfYestJ}_1+2\norm{\qYJ-\qfYestJ}_2\norm{\qfYestJ-\GJ}_2\\ \numberthis \label{eq:up1}
  &= 2-2\norm{\qfYestJ}_1+ \underbrace{ \norm{\qfYestJ}_2^2- \norm{\qYJ}_2^2}_{\text{(I)}}+\norm{\qYJ-\qfYestJ}^2_2 +2\norm{\qYJ-\qfYestJ}_2\underbrace{\norm{\qfYestJ-\GJ}_2}_{\text{(II)}}. 
\end{align*}
In what follows, we bound the terms denoted by (I) and (II).
\paragraph{Bounding (I):} From the Minkowski's inequality for $2$-norm, we have
\begin{align*}
\norm{\qfYestJ}_2^2 &\leq \Big(\norm{\qYJ}_2+\norm{\qfYestJ-\qYJ}_2  \Big)^2\\
&=\norm{\qYJ}^2_2+\norm{\qfYestJ-\qYJ}^2_2+2\norm{\qYJ}_2 \norm{\qfYestJ-\qYJ}_2\\
&\leq \norm{\qYJ}^2_2+{\norm{\qfYestJ-\qYJ}^2_2+2 \norm{\qfYestJ-\qYJ}_2},
\end{align*}
where the second inequality is due Bessel's inequality and the following chain of inequalities $$\norm{\qYJ}_2\leq \norm{\qfY}_2= \norm{\rho_{XY}(\id\tensor \sigma^3)}_2=\norm{\rho_{XY}}_2\leq \norm{\rho_{XY}}_1= 1,$$
where we used the fact that $\qfY=-\sqrt{\rho_{XY}}(\id \tensor \sigma^3)\sqrt{\rho_{XY}}$ which also equals to $\rho_{XY}(\id\tensor \sigma^3)$, and that $\norm{\cdot}_2\leq \norm{\cdot}_1$.
 Hence, the term (I) in \eqref{eq:up1} is upper bounded as
\begin{align}\label{eq:bound on (I)}
\text{(I)}\leq \lambda_1\deq {\norm{\qfYestJ-\qYJ}^2_2+2 \norm{\qfYestJ-\qYJ}_2}.
\end{align}

\paragraph{Bounding (II):} From \eqref{eq:fbar-sign}, we have
\begin{align*}
\norm{\qfYestJ-\GJ}_2^2 &= 1+\norm{\qfYestJ}_2^2-2\norm{\qfYestJ}_1\\
&\stackrel{(a)}{\leq} 1+ 2( \norm{\qYJ}^2_2+ \norm{\qYJ-\qfYestJ}^2_2)-2\norm{\qfYestJ}_1\\
&\stackrel{(b)}{=} 1+ 2( \norm{\qYJ}^2_2+ \norm{\qYJ-\qfYestJ}^2_2)-2\big(\norm{\qYJ}_1+ ( \norm{\qfYestJ}_1-\norm{\qYJ}_1)\big)\\
&= 1+ 2( \norm{\qYJ}^2_2- \norm{\qYJ}_1)+ 2 \norm{\qYJ-\qfYestJ}^2_2-2\big( \norm{\qfYestJ}_1-\norm{\qYJ}_1\big)\\
&\stackrel{(c)}{\leq}  1+ 2 \norm{\qYJ-\qfYestJ}^2_2-2\big( \norm{\qfYestJ}_1-\norm{\qYJ}_1\big)\\\numberthis \label{eq:temp1}
&\stackrel{(d)}{\leq } 1+ 2 \norm{\qYJ-\qfYestJ}^2_2+2\sqrt{2^d}\norm{\qYJ-\qfYestJ}_2,
\end{align*}
where $(a)$ follows from the Minkowski's inequality for $2$-norm and the inequality $(x+y)^2\leq 2(x^2+y^2)$. Equality $(b)$ follows by adding and subtracting  $\norm{\qYJ}_1$. Inequality $(c)$ holds from the inequality $\norm{\cdot}_2\leq \norm{\cdot}_1$ and the fact that $\norm{\qYJ}_2 \leq 1$ which implies that $\norm{\qYJ}^2_2\leq \norm{\qYJ}_2\leq \norm{\qYJ}_1$. 
Lastly, inequality $(d)$ holds because of the following chain of inequalities
\begin{align}\label{eq:normdiff_leq_eps}
\Big|\norm{\qYJ}_1-\norm{\qfYestJ}_1\Big| \leq \norm{\qYJ-\qfYestJ}_1\leq \sqrt{2^d}\norm{\qYJ-\qfYestJ}_2,
\end{align}
where the first inequality is due to the Minkowski's inequality for $1$-norm and 
the second inequality is due to the inequality $\norm{\cdot}_1\leq \sqrt{dim} \norm{\cdot}_2$ and the fact that $dim=2^d$. 

Next, we show that the quantity $\big\| \qfYestJ-\GJ\big\|_2$ without the square is upper bounded by the same term as in the right-hand side of \eqref{eq:temp1}. That is 
\begin{align}\label{eq:bound on (II)}
\text{(II)} &= \big\|\qfYestJ-\GJ\big\|_2\leq \lambda_2\deq 1+ 2 \norm{\qYJ-\qfYestJ}^2_2+2\sqrt{2^d}\norm{\qYJ-\qfYestJ}_2.
\end{align}
The argument is as follows: if $\big\|\qfYestJ-\GJ\big\|_2 \leq 1$, then the upper bound holds trivially as $\lambda_2\geq 1$; 
otherwise, if $\big\|\qfYestJ-\GJ\big\|_2 > 1$, then quantity is less than its squared, i.e.,  $\big\|\qfYestJ-\GJ\big\|_2 \leq \big\|\qfYestJ-\GJ\big\|_2^2$.   In that case, we obtain an upper bound using \eqref{eq:temp1}. 

As a result of the bounds in \eqref{eq:up1}, \eqref{eq:bound on (I)}, and \eqref{eq:bound on (II)} we obtain that 
\begin{align*}
4\Loss(\mathcal{M}_{\mathcal{J}})  &\leq 2-2\norm{\qfYestJ}_1+ \lambda_1+\norm{\qYJ-\qfYestJ}^2_2+  2 \lambda_2 \norm{\qYJ-\qfYestJ}_2\\
&= 2-2\norm{\qYJ}_1+2\Big(\norm{\qYJ}_1-\norm{\qfYestJ}_1\Big)  +\lambda_1+\norm{\qYJ-\qfYestJ}^2_2+ 2 \lambda_2 \norm{\qYJ-\qfYestJ}_2\\
&\leq 2-2\norm{\qYJ}_1+2\norm{\qYJ-\qfYestJ}_2+ \lambda_1+\norm{\qYJ-\qfYestJ}^2_2+  2 \lambda_2 \norm{\qYJ-\qfYestJ}_2,
\end{align*}
where the last inequality is due to \eqref{eq:normdiff_leq_eps}.  Next, from the definition of $\lambda_1$ and $\lambda_2$, the right hand side of the above inequality simplifies to the following
\begin{align*}
4\Loss(\mathcal{M}_{\mathcal{J}}) &\leq 2-2\norm{\qYJ}_1+ 4\norm{\qYJ-\qfYestJ}^3_2+ 2\norm{\qYJ-\qfYestJ}^2_2+4\sqrt{2^d}\norm{\qYJ-\qfYestJ}^2_2+  6 \norm{\qYJ-\qfYestJ}_2,
\end{align*}
We, further upper bound the right hand side by replacing  $\norm{\qYJ-\qfYestJ}_2$ with $\sqrt{2^d}\norm{\qYJ-\qfYestJ}_2$ in the third, fourth and sixth terms above. As a result we get 
\begin{align*}
4\Loss(\mathcal{M}_{\mathcal{J}}) \leq 2-2\norm{\qYJ}_1+4U(\sqrt{2^d}\norm{\qYJ-\qfYestJ}_2),
\end{align*}
where $U(x) = x^3+\frac{3}{2}x^2+\frac{3}{2}x$ as in the statement of the lemma. Dividing both sides by $4$ completes the proof.

\end{document}